%% file: sample-sigconf.tex
\documentclass[sigconf]{acmart}

\renewcommand\footnotetextcopyrightpermission[1]{} % removes footnote with conference information in first column
\usepackage{booktabs} % For formal tables
\usepackage{url}
\usepackage{mathrsfs}
\usepackage[mathscr]{eucal}
\usepackage[linesnumbered,ruled,vlined]{algorithm2e}
\usepackage{multirow}
\usepackage{subcaption}
\usepackage{amsthm}
\usepackage{amsmath}

\usepackage{tikz}
\usetikzlibrary{arrows,positioning,automata,calc,shapes}
\usepackage{pgfplots}
\pgfplotsset{compat=newest, scaled z ticks=false} 
\pgfplotsset{plot coordinates/math parser=false}
\newlength\figureheight 
 \newlength\figurewidth

\newcommand{\squishlist}{
    \begin{list}{$\bullet$}
    { \setlength{\itemsep}{0pt}
        \setlength{\parsep}{1pt}
        \setlength{\topsep}{1pt}
        \setlength{\partopsep}{0pt}
        \setlength{\leftmargin}{1em} %1.5em
        \setlength{\labelwidth}{1em}
        \setlength{\labelsep}{0.5em}
    						 } }

\newcommand{\squishlisttwo}{
    \begin{list}{$\bullet$}
        { \setlength{\itemsep}{0pt}
            \setlength{\parsep}{0pt}
            \setlength{\topsep}{0pt}
            \setlength{\partopsep}{0pt}
            \setlength{\leftmargin}{2em}
            \setlength{\labelwidth}{1.5em}
            \setlength{\labelsep}{0.5em} } }

\newcommand{\squishend}{
    \end{list}  }

\begin{document}
\title{Fairness-aware Personalized Ranking Recommendation via Adversarial Learning}

\author{Ziwei Zhu}
\affiliation{
 \institution{Texas A\&M University}}
\email{zhuziwei@tamu.edu}

\author{Jianling Wang}
\affiliation{
 \institution{Texas A\&M University}}
\email{jlwang@tamu.edu}

\author{James Caverlee}
\affiliation{
 \institution{Texas A\&M University}}
\email{caverlee@tamu.edu}

% \maketitle

\fancyhead{}

\begin{abstract}
Recommendation algorithms typically build models based on historical user-item interactions (e.g., clicks, likes, or ratings) to provide a personalized ranked list of items. These interactions are often distributed unevenly over different groups of items due to varying user preferences. However, we show that recommendation algorithms can inherit or even amplify this imbalanced distribution, leading to unfair recommendations to item groups. Concretely, we formalize the concepts of ranking-based statistical parity and equal opportunity as two measures of fairness in personalized ranking recommendation for item groups. Then, we empirically show that one of the most widely adopted algorithms -- Bayesian Personalized Ranking -- produces unfair recommendations, which motivates our effort to propose the novel fairness-aware personalized ranking model. The debiased model is able to improve the two proposed fairness metrics while preserving recommendation performance. Experiments on three public datasets show strong fairness improvement of the proposed model versus state-of-the-art alternatives.

This is paper is an extended and reorganized version of our SIGIR 2020~\cite{zhu2020measuring} paper. In this paper, we re-frame the studied problem as `item recommendation fairness' in personalized ranking recommendation systems, and provide more details about the training process of the proposed model and details of experiment setup.

\end{abstract}

\maketitle

\input{samplebody-conf}

\bibliographystyle{ACM-Reference-Format}
\bibliography{sample-bibliography}

\clearpage
\appendix

\section{Model Training}
\label{sec:model_training}

\setlength{\textfloatsep}{1pt}
\begin{algorithm}[t!]
 \KwIn{Training data $\mathcal{D}$, adversarial regularizer $\alpha$, KL-loss regularizer $\beta$, $L_2$ regularizer $\lambda_\mathbf{\Theta}$, and learning rate for BPR $\eta_{BPR}$, learning rate for the adversary $\eta_{Adv}$ \;}
 \KwOut{BPR parameters $\mathbf{\Theta}$\;}
Randomly Initialize model parameters $\mathbf{\Theta}$ for BPR, and $\mathbf{\Psi}$ for the adversary\;
\Repeat{\textbf{converge}}{
 \For{$(u^\prime,i^\prime,j^\prime)$ in $\mathcal{D}$}{
    Update $\mathbf{\Psi}$ based on $\pmb{\mathscr{L}}_{Adv}(i^\prime)$ by backpropagation\;
    Update $\mathbf{\Psi}$ based on $\pmb{\mathscr{L}}_{Adv}(j^\prime)$ by backpropagation\;
}
Randomly draw a mini-batch $\mathcal{D}^{mini}$ from $\mathcal{D}$\;
\For{$(u,i,j)$ in $\mathcal{D}^{mini}$}{
    
    Update $\mathbf{\Theta}$ based on gradient $\frac{\partial(\pmb{\mathscr{L}}_{BPR}(u,i,j)+\alpha(\pmb{\mathscr{L}}_{Adv}(i)+\pmb{\mathscr{L}}_{Adv}(j))+\beta\pmb{\mathscr{L}}_{KL})}{\partial\mathbf{\Theta}}$\;
}
}
Return $\mathbf{\Theta}$\;
\caption{Training algorithm for DPR-RSP.}
 \label{alg:training}
%  \vspace{0.cm}
\end{algorithm}

The model training process for DPR-RSP can be summarized in Algorithm~\ref{alg:training}, where we train the model in a mini-batch manner. In each epoch, there are two phases: we first update the weights in the MLP adversary to maximize the classification objective, then update the BPR to minimize pairwise ranking loss, classification objective and KL-loss all together. Following the adversarial training proposed in \cite{louppe2017learning}, in each iteration, we first update the MLP adversary by the whole dataset, then update BPR by a mini-batch, which empirically leads to  faster convergence. And in practice, we usually first pre-train the BPR model for several epochs and then add in the adversarial training part.

Algorithm~\ref{alg:training} can be easily extended to the DPR-REO model by two minor modifications: first remove the negative samples update steps for MLP adversary (line 5); then replace the update rule of line 8 with the gradient $\frac{\partial((\pmb{\mathscr{L}}_{BPR}(u,i,j)+\alpha\pmb{\mathscr{L}}_{Adv}(i))+\beta\pmb{\mathscr{L}}_{KL})}{\partial\mathbf{\Theta}}$.

\section{Objective Functions of Baselines}
\label{sec:baseline_objective_function}

We replace the original Sum of Squared Error (SSE) in FATR~\cite{zhu2018fairness} with BPR pairwise loss function introduced in Equation~\ref{equ:BPR}, and add the proposed KL-loss introduced in Section~\ref{sec:KL-loss} to bridge the predicted score to the rankings. Hence, We have the new FATR:
\begin{equation*}
\begin{aligned}
\centering
&\underset{\mathbf{\Theta}=\{\mathbf{P},\widetilde{\mathbf{Q}}\}}{\text{min}}
&&\pmb{\mathscr{L}}=  \pmb{\mathscr{L}}_{BPR}+\dfrac{\lambda}{2}\lVert{\mathbf{Q}^{\prime\prime}}^\top\mathbf{Q}^\prime\rVert^2_{\text{F}}+\dfrac{\gamma}{2}\lVert\mathbf{\Theta}\rVert^2_{\textrm{F}}+\beta\pmb{\mathscr{L}}_{KL},\\
&\text{s.t.}
&&\widetilde{\mathbf{Q}}=[\mathbf{Q}^\prime; \mathbf{Q}^{\prime\prime}],
\end{aligned}
\label{equ:FATR}
\end{equation*}
where $\mathbf{P}\in\mathbb{R}^{d\times N}$ is the latent factor matrix for user, $\widetilde{\mathbf{Q}}\in\mathbb{R}^{d\times M}$ is the latent factor matrix for item with last $A$ dimensions indicating the groups, $\mathbf{Q}^\prime\in\mathbb{R}^{d-A\times M}$ is the first $d-A$ dimensions to be learned by training, $\mathbf{Q}^{\prime\prime}\in\mathbb{R}^{A\times M}$ is the sensitive dimensions indicating groups, and $[\mathbf{Q}^\prime; \mathbf{Q}^{\prime\prime}]$ concatenates $\mathbf{Q}^\prime$ and $\mathbf{Q}^{\prime\prime}$ vertically.

In the similar way, we also modify the original regularization-based methods by adopting BPR pairwise loss and KL-loss to enhance fairness of RSP and REO. We have the model Reg-RSP:
\begin{equation}
\begin{aligned}
\centering
\underset{\mathbf{\Theta}=\{\mathbf{P},\mathbf{Q}\}}{\text{min}}\,\,
\pmb{\mathscr{L}}=\pmb{\mathscr{L}}_{BPR}+\dfrac{\lambda}{2}({\overline{s}_{g_1}-\overline{s}_{g_2}})^2+\dfrac{\gamma}{2}\lVert\mathbf{\Theta}\rVert^2_{\textrm{F}}+\beta\pmb{\mathscr{L}}_{KL},
\end{aligned}
\label{equ:Reg-SP}
\end{equation}
where $\overline{s}_{g_1}$ and $\overline{s}_{g_2}$ calculates the average scores for groups $g_1$ and $g_2$ accordingly. Similarly, we can have Reg-REO to augment REO:
\begin{equation}
\begin{aligned}
\centering
\underset{\mathbf{\Theta}=\{\mathbf{P},\mathbf{Q}\}}{\text{min}}\,\,
\pmb{\mathscr{L}}=\pmb{\mathscr{L}}_{BPR}+\dfrac{\lambda}{2}({\overline{s}_{g_1|y=1}-\overline{s}_{g_2|y=1}})^2+\dfrac{\gamma}{2}\lVert\mathbf{\Theta}\rVert^2_{\textrm{F}}+\beta\pmb{\mathscr{L}}_{KL},
\end{aligned}
\label{equ:Reg-EO}
\end{equation}
where $\overline{s}_{g_1|y=1}$ and $\overline{s}_{g_2|y=1}$ calculates the average scores of  user-item pairs with positive feedback for groups $g_1$ and $g_2$ accordingly.

\end{document}

%% file: samplebody-conf.tex
\section{Introduction}

The social and ethical concerns raised by recommenders are increasingly attracting attention, including issues like filter bubbles~\cite{nguyen2014exploring}, transparency~\cite{nilashi2016recommendation}, and accountability~\cite{xavier2016learning}. In particular, \textit{item recommendation unfairness} -- wherein one or more groups of items are systematically under-recommended -- is one of the most common but harmful issues in a personalized ranking recommender. Unfairness is common in scenarios where the training data used to learn a recommender has an imbalanced distribution of feedback for different item groups due to the inherent uneven preference distribution in the real world~\cite{yao2017beyond}. For example, ads for non-profit jobs may be clicked at a lower rate than high-paying jobs, so that a recommendation model trained over this skewed data will inherit or even amplify this imbalanced distribution. This can result in ads for non-profit jobs being unfairly under-recommended.

Previous works on recommendation fairness~\cite{yao2017beyond,zhu2018fairness,kamishima2013efficiency,kamishima2016model,kamishima2017considerations,kamishima2018recommendation} mainly focus on investigating how to produce similar predicted score distributions for different groups of items (in other words, by removing the influence of group information when predicting preference scores). The main drawback of these works is that they mainly focus on the perspective of \textit{predicted preference scores}~\cite{yao2017beyond,kamishima2013efficiency,kamishima2016model,kamishima2018recommendation,kamishima2017considerations,zhu2018fairness}. In practice, however, predicted scores are an intermediate step towards a ranked list of items that serves as the final recommendation result, and having similar predicted scores does not necessarily lead to a fair ranking result, suggesting the importance of directly measuring fairness over rankings instead of scores. 

\begin{figure}[t!]
    % \vspace{-14pt}
    \centering
   \includegraphics[ width=0.94\linewidth ]{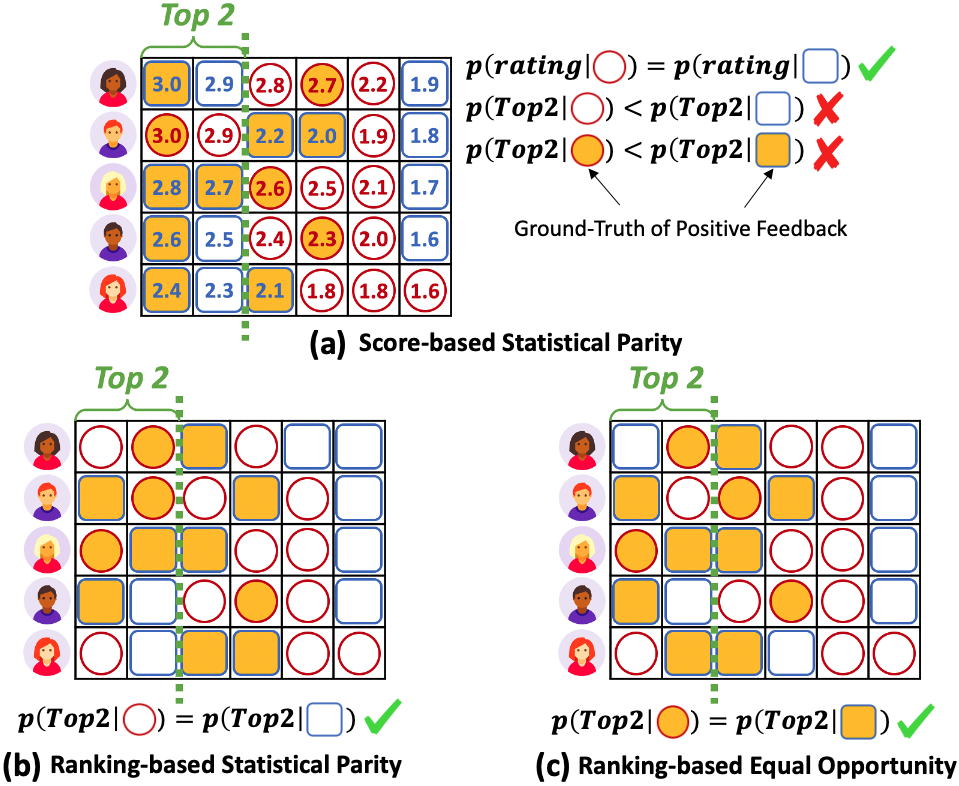} 
   \vspace{-10pt}
    \caption{(a) is an example following score-based statistical parity from previous works~\cite{kamishima2017considerations,zhu2018fairness}. (b) and (c) are examples of the proposed metrics: (b) is ranking-based statistical parity, and (c) is ranking-based equal opportunity.}
    \vspace{-12pt}
    \label{fig:example} 
\end{figure}

Take the most frequently adopted concept -- statistical parity (or demographic parity)~\cite{yao2017beyond,zhu2018fairness,kamishima2017considerations} -- as an example. Fairness of statistical parity encourages identical predicted score distributions for different groups. Figure~\ref{fig:example}a shows an example following the statistical parity constraint, where rows represent users, and blue squares and red circles represent two different groups of items. Each group has three items. The numbers in the matrix are the predicted scores from users to items. The items are ranked by the scores in descending order for each user. The top-2 items are recommended to users, and the yellow background in some circle and square items represents positive ground-truth. In this example, the blue square group and red circle group have exactly the same score distribution (a uniform distribution over the range $[1.6, 3.0]$). However, the recommendation result is unfair for the red circle group, which only has two items being recommended. Therefore, we propose two new fairness metrics calculated directly over the ranking results --  \textit{ranking-based statistical parity (RSP)} and \textit{ranking-based equal opportunity (REO)}. 

\smallskip
\noindent\textbf{Motivating Example: Ranking-Based Statistical Parity.} Unlike traditional score-based statistical parity, RSP encourages the probabilities for different groups being recommended (being ranked within top $k$) to be similar. Figure~\ref{fig:example}b provides an example following RSP, where the probabilities of being ranked within the top-2 for the two groups are both $\frac{1}{3}$. However, while RSP encourages a fair ranking result, it neglects the intrinsic group quality or user preference imbalance of the data, which may exert unfairness to the popular group. For instance, in Figure~\ref{fig:example}a and Figure~\ref{fig:example}b, the yellow background in some entries represents that the marked item has the ground-truth that it is liked by the corresponding user (i.e. it forms a matched user-item pair). Then, we see that the blue square group is more preferred by users than the red circle group since the numbers of positive user-item pairs in the two groups are $8:4$. Hence, keeping the same recommendation probability in Figure~\ref{fig:example}b is unfair for the blue square group.

\smallskip
\noindent\textbf{Motivating Example: Ranking-Based Equal Opportunity.} Taking into consideration the ground-truth of user-item matching, we propose the metric REO that encourages the probabilities of being correctly recommended to matched users to be the same across item groups. A positive example is given in Figure~\ref{fig:example}c, where the probability of being ranked in top-2 to matched users is $\frac{1}{2}$ for both item groups. Figure~\ref{fig:example}a and Figure~\ref{fig:example}b show the negative examples, where in Figure~\ref{fig:example}a the ranking result shows favor to the blue square group, while Figure~\ref{fig:example}b is biased towards the red circle group. 

Due to their distinct characteristics, RSP and REO are applicable for different scenarios. RSP-based fairness is usually important and needs to be enhanced for situations where there are sensitive attributes attached to the subjects being recommended, such as considering fairness across genders or races of job candidates when they are recommended to companies for hiring. Because in this case, social ethics require equal exposure probability for different demographic groups in the system. While REO-based fairness applies to more general scenarios without sensitive attribute, like movies or songs recommendation. Unfairness regarding REO in these systems can potentially exerts damaging influence to both users and item providers. On the one hand, user needs corresponding to minority item groups are not fully acknowledged, leading to lower user satisfaction. On the other hand, item providers of minority groups may not receive enough exposure hurting their economic gains. 

\smallskip
\noindent\textbf{Contributions.}
With these two metrics in mind, we empirically demonstrate that a fundamental recommendation model -- Bayesian Personalized Ranking (BPR)~\cite{rendle2009bpr} -- is vulnerable to unfairness, which motivates our efforts to address it. Then, we show how to improve the fairness based on the two introduced metrics through a debiased personalized ranking model (DPR) that has two key features: a multi-layer perceptron adversary that seeks to enhance the score distribution similarity among item groups and a KL Divergence based regularization term that aims to normalize the score distribution for each user. Incorporating these two components together, RSP (or REO depending on how we implement the adversary learning) based fairness can be significantly improved while preserving recommendation quality at the same time. Extensive experiments on three public datasets show the effectiveness of the proposed model over state-of-the-art alternatives. In general, DPR is able to reduce the unfairness corresponding to the two metrics for BPR by $67.3\%$ on average (with an improvement of $48.7\%$ over the best baseline), while only decreasing $F1@15$ versus BPR by $4.1\%$ on average (with an improvement of $16.4\%$ over the best baseline).

\section{Related Work}

\noindent\textbf{Recommendation Fairness.}
Many efforts have focused on improving fairness for recommendation tasks in the context of explicit rating prediction. More specifically, from the perspective of eliminating unfairness based on statistical parity, Kamishima et al. proposed regularization-based models~\cite{kamishima2013efficiency,kamishima2018recommendation} to penalize unfairness in the predicted ratings; and Yao et al.~\cite{yao2017beyond} proposed three fairness metrics on the user side in rating prediction tasks and generalized the regularization-based model to balance recommendation quality and the proposed metrics. Recent works have started to shift attention toward recommendation fairness in personalized ranking tasks. However unlike this paper, most do not align fairness with the ranking results. For example, although under the ranking recommendation setting, Kamishima et al.~\cite{kamishima2017considerations} adopted a regularization-based approach to enhance fairness for predicted preference scores; and Zhu et al.~\cite{zhu2018fairness} proposed a parity-based model over predicted scores by first isolating sensitive features and then extracting the sensitive information. Research in \cite{burke2018balanced}, \cite{Fairness2019Geyik} and \cite{beutel2019fairness} proposed metrics that consider the ranking results. However, there are three main differences between them and the present work: i) neither of \cite{burke2018balanced} nor \cite{Fairness2019Geyik} takes ground truth of user-item matching into consideration; ii) \cite{burke2018balanced} and \cite{beutel2019fairness} only consider two-group scenarios; and iii) \cite{beutel2019fairness} and \cite{Fairness2019Geyik} consider fairness among item groups for individual users rather than consider system-level fairness for different item groups. In sum, the main differences between this work and previous works are that the fairness we investigate is over ranking results among multiple groups; is based on both statistical parity and equal opportunity; and is calculated from the system-level.

\smallskip
\noindent\textbf{Other Related Topics.}
There are some recent efforts investigating topics related to recommendation fairness. For example, Beutel et al.~\cite{beutel2017beyond} and Krishnan et al.~\cite{krishnan2018adversarial} explored approaches to improve the fairness w.r.t. recommendation accuracy for niche items. Recommendation diversity~\cite{hurley2011novelty}, which requires as many groups as possible appearing in the recommendation list for each user, is related to the metric RSP in this work, but fundamentally different. Another similar concept to RSP called calibrated recommendations is proposed by Steck~\cite{steck2018calibrated}, which encourages the same group proportions as the historical record for each user and can be regarded as a special fairness for individual users. The main differences of our work and these previous works are that research of recommendation diversity and recommendation calibration investigate the distribution skews for each individual user rather than for the whole system, and they only consider the recommendation distributions without taking into account the ground truth of user preference and item quality as in this work.

\section{Fairness in Personalized Ranking}
In this section, we first describe the personalized ranking problem and ground our discussion through a treatment of Bayesian Personalized Ranking (BPR). Next, we introduce two proposed fairness metrics for personalized ranking. Last, we empirically demonstrate that BPR is vulnerable to data bias and tends to produce unfair recommendations. 

\subsection{Bayesian Personalized Ranking}
\label{sec:BPR}

Given $N$ users $\mathcal{U} = \{1, 2, \ldots, N\}$ and $M$ items $\mathcal{I} = \{1, 2, \ldots, M\}$, the personalized ranking problem is to recommend a list of $k$ items to each user $u$ based on the user's historical behaviors $\mathcal{I}^+_{u} = \{i, j,\ldots \}$, where $i, j,\ldots$ are the items $u$ interacts with before (and so can be regarded as implicit positive feedback). Bayesian Personalized Ranking (BPR)~\cite{rendle2009bpr} is one of the most influential methods to solve this problem, which is the foundation of many cutting edge personalized ranking algorithms (e.g. \cite{he2016vbpr, he2018adversarial}). BPR adopts matrix factorization~\cite{koren2009matrix} as the base and minimizes a pairwise ranking loss, formalized as:
% \vspace{-5pt}
\begin{equation}
% \vspace{-5pt}
\begin{aligned}
\centering
 \underset{\mathbf{\Theta}}{\text{min}}\,\, \mathscr{L}_{BPR}=-\sum_{u\in\mathcal{U}}\sum_{\substack{i\in\mathcal{I}_{u}^+\\j\in\mathcal{I}{\setminus}\mathcal{I}_u^+}}ln\, \sigma(\widehat{y}_{u,i}-\widehat{y}_{u,j}) + \dfrac{\lambda_{\mathbf{\Theta}}}{2}\lVert\mathbf{\Theta}\rVert^2_{\text{F}},
\end{aligned}
\label{equ:BPR}
\end{equation}
where $\widehat{y}_{u,i}$ and $\widehat{y}_{u,j}$ are the predicted preference scores calculated by the matrix factorization model for user $u$ to positive item $i$ and sampled negative item $j$; $\sigma(\cdot)$ is the Sigmoid function; $\lVert \cdot \rVert_{\text{F}}$ is the Frobenius norm; $\mathbf{\Theta}$ represents the model parameters, i.e., $\mathbf{\Theta}=\{\mathbf{P},\mathbf{Q}\}$, where $\mathbf{P}$ and $\mathbf{Q}$ are the latent factor matrices for users and items; and $\lambda_{\mathbf{\Theta}}$ is the trade-off weight for the l2 regularization.

With the trained BPR, we can predict the preference scores toward all un-interacted items and rank them in descending order for user $u$. A list of items with the top $k$ largest scores $\{R_{u,1},R_{u,2}, \ldots, R_{u,k}\}$ will be recommended to user $u$, where $R_{u,k}$ is the item id at the ranked $k$ position.

\subsection{Fairness Metrics for Personalized Ranking}
\label{sec:metrics}
However, there is no notion of fairness in such a personalized ranking model. Here, we assume a set of $A$ sensitive groups $\mathcal{G}=\{g_1, g_2, \ldots,g_A\}$, and every item in $\mathcal{I}$ belongs to one or more groups. A group here could correspond to gender, ethnicity, or other item attributes. We define a function $G_{g_a}(i)$ to identify whether item $i$ belongs to group $g_a$, if it does, the function returns $1$, otherwise $0$. The high-level goal of a fairness-aware recommender is to enhance the fairness among different groups such that no group is under-recommended. 

Previous definitions of recommendation fairness ~\cite{zhu2018fairness,kamishima2017considerations,burke2018balanced} have two main drawbacks: (i) the fairness metrics are calculated over the predicted scores, which are not aligned with the ranking results; and (ii) the definitions are mainly based on the concept of statistical parity, which does not take the ground-truth of user-item matching into account. Therefore, we propose two new fairness metrics over the ranking of items from different groups.

\smallskip
\noindent\textbf{Ranking-based Statistical Parity (RSP).}
Statistical parity requires the probability distributions of model outputs for different input groups to be the same. In a similar way, for the personalized ranking task, statistical parity can be defined as forcing the ranking probability distributions of different item groups to be the same. Because conventionally only the top-$k$ items will be recommended to users, we focus on the probabilities of being ranked in top-$k$, which is also aligned with basic recommendation quality evaluation metrics such as \textit{precision@k} and \textit{recall@k}. As a result, we propose the ranking-based statistical parity metric -- RSP, which encourages $P(R@k|g=g_1)=P(R@k|g=g_2)=\ldots=P(R@k|g=g_A)$, where $R@k$ represents `being ranked in top-$k$', and $P(R@k|g=g_a)$ is the probability of items in group $g_a$ being ranked in top-$k$. Formally, we calculate the probability as follows:
\begin{equation*}
\begin{aligned}
\centering
 P(R@k|g=g_a) = \frac{\sum_{u=1}^{N}\sum_{i=1}^{k}G_{g_a}(R_{u,i})}{\sum_{u=1}^{N}\sum_{i\in{\mathcal{I}{\setminus}\mathcal{I}_u^+}}G_{g_a}(i)},
\end{aligned}
\label{equ:SP_Prob}
\end{equation*}
where $\sum_{i=1}^{k}G_{g_a}(R_{u,i})$ calculates how many un-interacted items from group $g_a$ are ranked in top-$k$ for user $u$, and $\sum_{i\in{\mathcal{I}{\setminus}\mathcal{I}_u^+}}G_{g_a}(i)$ calculates how many un-interacted items belong to group $g_a$ for $u$. Last, we compute the \textit{relative standard deviation} (to keep the same scale for different $k$) over the probabilities to determine $RSP@k$:
\begin{equation*}
\begin{aligned}
\centering
 RSP@k = \frac{std(P(R@k|g=g_1),\ldots,P(R@k|g=g_A))}{mean(P(R@k|g=g_1),\ldots,P(R@k|g=g_A))},
\end{aligned}
\label{equ:SP}
\end{equation*}
where $std(\cdot)$ calculates the standard deviation, and $mean(\cdot)$ calculates the mean value.

\smallskip
\noindent\textbf{Ranking-based Equal Opportunity (REO).} Our second metric is based on the concept of equal opportunity~\cite{beutel2017data,hardt2016equality,zhang2018mitigating}, which encourages the true positive rates (TPR) of different groups to be the same. Take a binary classification task with two groups as an example, equal opportunity requires:
\begin{equation*}
\begin{aligned}
\centering
 P(\widehat{c}=1|g=0,c=1)=P(\widehat{c}=1|g=1,c=1),
\end{aligned}
\label{equ:EO_classification}
\end{equation*}
where $c$ is the ground-truth label, $\widehat{c}$ is the predicted label; $P(\widehat{c}=1|g=0,c=1)$ represents the TPR for group 0, $P(\widehat{c}=1|g=1,c=1)$ is the TPR for group 1. Similarly, in the personalized ranking system, equal opportunity demands the ranking based TPR for different groups to be the same. We can define the TPR as the probability of being ranked in top-$k$ given the ground-truth that the user likes the item, noted as $P(R@k|g=g_a,y=1)$, where $y=1$ represents items are liked by users. The probability can be calculated by:
\begin{equation*}
\begin{aligned}
\centering
 P(R@k|g=g_a,y=1) = \frac{\sum_{u=1}^{N}\sum_{i=1}^{k}G_{g_a}(R_{u,i})Y(u,R_{u,i})}{\sum_{u=1}^{N}\sum_{i\in{\mathcal{I}{\setminus}\mathcal{I}_u^+}}G_{g_a}(i)Y(u,i)},
\end{aligned}
\label{equ:EO_Prob}
\end{equation*}
where $Y(u,R_{u,i})$ identifies the ground-truth label of a user-item pair $(u,R_{u,i})$, if item $R_{u,i}$ is liked by user $u$, returns 1, otherwise 0 (in practice, $Y(u,i)$ identifies whether a user-item pair $(u,i)$ is in the test set); $\sum_{i=1}^{k}G_{g_a}(R_{u,i})Y(u,R_{u,i})$ counts how many items in test set from group $g_a$ are ranked in top-$k$ for user $u$, and $\sum_{i\in{\mathcal{I}{\setminus}\mathcal{I}_u^+}}G_{g_a}(i)Y(u,i)$ counts the total number of items from group $g_a$ in test set for user $u$. Similar to RSP, we calculate the relative standard deviation to determine $REO@k$:
\begin{equation*}
\begin{aligned}
\centering
 REO@k=\frac{std(P(R@k|g=g_1,y=1)\ldots P(R@k|g=g_A,y=1))}{mean(P(R@k|g=g_1,y=1)\ldots P(R@k|g=g_A,y=1))}.
\end{aligned}
\label{equ:EO}
\end{equation*}

For classification tasks, TPR is the recall of classification, and for personalized ranking, the probability $P(R@k|g=g_a,y=1)$ is \textit{recall@k} of group $g_a$. In other words, mitigating REO-based bias requires $recall@k$ for different groups to be similar.

Note that for both $RSP@k$ and $REO@k$, \textbf{lower values indicate the recommendations are fairer}. In practice, RSP is particularly important in scenarios where humans or items with sensitive attributes are recommended (such as political news). Because RSP-based fairness in these scenarios leads to social issues like gender discrimination during recruiting or political ideology unfairness during election campaigns. Conversely, REO-based fairness is supposed to be enhanced in general item recommendation systems so that no user need is ignored, and all items have the chance to be exposed to matched users who like them.

\subsection{BPR is Vulnerable to Unfairness}
\label{sec:BPR_Experiment}

\begin{table}[t!]\small
 %  \vspace{-10pt}
  \begin{center}
  \setlength{\tabcolsep}{2.7pt}
  \renewcommand{\arraystretch}{.8}
\begin{tabular}{c|c|ccc}
\hline\hline
                        & Group        & \#Item & \#Feedback & $\frac{\#feedback}{\#item}$ \\ \hline
\multirow{7}{*}{ML1M}   & Sci-Fi        & 271    & 157,290    & 580.41                      \\
                        & Adventure     & 276    & 133,946    & 485.31                      \\
                        & Crime         & 193    & 79,528     & 412.06                      \\
                        & Romance       & 447    & 147,501    & 329.98                      \\
                        & Children's    & 248    & 72,184     & 291.06                      \\
                        & Horror        & 330    & 76,370     & 231.42                      \\ \cline{2-5} 
                        & Relative std  & -      & -          & \textbf{0.33}                        \\ \hline
\multirow{5}{*}{Yelp}   & American(New) & 1610   & 91,519     & 56.84                       \\
                        & Japanese      & 946    & 45,508     & 48.11                       \\
                        & Italian       & 1055   & 46,434     & 44.01                       \\
                        & Chinese       & 984    & 36,729     & 37.33                       \\ \cline{2-5} 
                        & Relative std  & -      & -          & \textbf{0.17}                        \\ \hline
\multirow{5}{*}{Amazon} & Grocery       & 749    & 49,646     & 66.28                       \\
                        & Office        & 892    & 37,776     & 42.35                       \\
                        & Pet           & 518    & 16,260     & 31.39                       \\
                        & Tool          & 606    & 14,771     & 24.37                       \\ \cline{2-5} 
                        & Relative std  & -      & -          & \textbf{0.44}                        \\ \hline\hline
\end{tabular}
  \end{center}
%   \vspace{.2cm}
        \caption{Group information in the three datasets.}
        \label{table:Datasets_groups} 
\vspace{-20pt}
\end{table}

\begin{table}[t!]\small
 %  \vspace{-10pt}
  \begin{center}
  \setlength{\tabcolsep}{1.5pt}
  \renewcommand{\arraystretch}{.9}
\begin{tabular}{cc|ccc|ccc}
\hline\hline
\multicolumn{2}{c|}{}                                                          & \multicolumn{3}{c|}{$P(R@k|g)$} & \multicolumn{3}{c}{$P(R@k|g,y=1)$} \\ \hline
\multicolumn{1}{c|}{}                        & Genres                         & @5       & @10      & @15     & @5        & @10       & @15       \\\hline
\multicolumn{1}{c|}{\multirow{7}{*}{ML1M}}   & Sci-Fi                         & .00654  & .01306  & .01949 & .09497   & .16819   & .22922   \\
\multicolumn{1}{c|}{}                        & Adventure                      & .00516  & .01022  & .01521 & .08884   & .15808   & .21657   \\
\multicolumn{1}{c|}{}                        & Crime                          & .00456  & .00888  & .01318 & .07469   & .13017   & .17941   \\
\multicolumn{1}{c|}{}                        & Romance                        & .00327  & .00665  & .01002 & .06448   & .12003   & .16366   \\
\multicolumn{1}{c|}{}                        & Children's                     & .00251  & .00494  & .00742 & .05852   & .10470   & .14464   \\
\multicolumn{1}{c|}{}                        & Horror                         & .00176  & .00354  & .00533 & .05399   & .10132   & .13985   \\ \cline{2-8} 
\multicolumn{1}{c|}{}                        & RSP or REO & \textbf{.41054}  & \textbf{.40878}  & \textbf{.40579} & \textbf{.20885}   & \textbf{.19316}   & \textbf{.18933}   \\ \hline
\multicolumn{1}{c|}{\multirow{5}{*}{Yelp}}   & American(New)                  & .00154  & .00302  & .00449 & .06345   & .10904   & .14497   \\
\multicolumn{1}{c|}{}                        & Japanese                       & .00111  & .00219  & .00328 & .04770   & .08207   & .11106   \\
\multicolumn{1}{c|}{}                        & Italian                        & .00093  & .00194  & .00297 & .03890   & .07087   & .09658   \\
\multicolumn{1}{c|}{}                        & Chinese                        & .00072  & .00146  & .00222 & .03376   & .05626   & .07961   \\ \cline{2-8} 
\multicolumn{1}{c|}{}                        & RSP or REO & \textbf{.28005}  & \textbf{.26376}  & \textbf{.25224} & \textbf{.24515}   & \textbf{.24290}   & \textbf{.22253}   \\ \hline
\multicolumn{1}{c|}{\multirow{5}{*}{Amazon}} & Grocery                        & .00283  & .00572  & .00869 & .03931   & .07051   & .09297   \\
\multicolumn{1}{c|}{}                        & Office                         & .00165  & .00336  & .00506 & .01196   & .02039   & .03180   \\
\multicolumn{1}{c|}{}                        & Pet                            & .00185  & .00348  & .00501 & .04815   & .07807   & .10215   \\
\multicolumn{1}{c|}{}                        & Tool                           & .00082  & .00165  & .00250 & .00552   & .01105   & .01519   \\ \cline{2-8} 
\multicolumn{1}{c|}{}                        &  RSP or REO & \textbf{.40008 } & \textbf{.40672}  & \textbf{.41549} & \textbf{.68285}   & \textbf{.65756}   & \textbf{.62175}   \\ \hline\hline
\end{tabular}
  \end{center}
%   \vspace{.2cm}
        \caption{Ranking probability distributions and RSP and REO metrics on three datasets by BPR. }
        \label{table:BPR_Vulnerable} 
\vspace{-24pt}
\end{table}

\begin{figure*}[t!]
    \vspace{-10pt}
    \centering
    \begin{subfigure}[t]{0.1725\textwidth}
        \centering
        \includegraphics[width=1\textwidth]{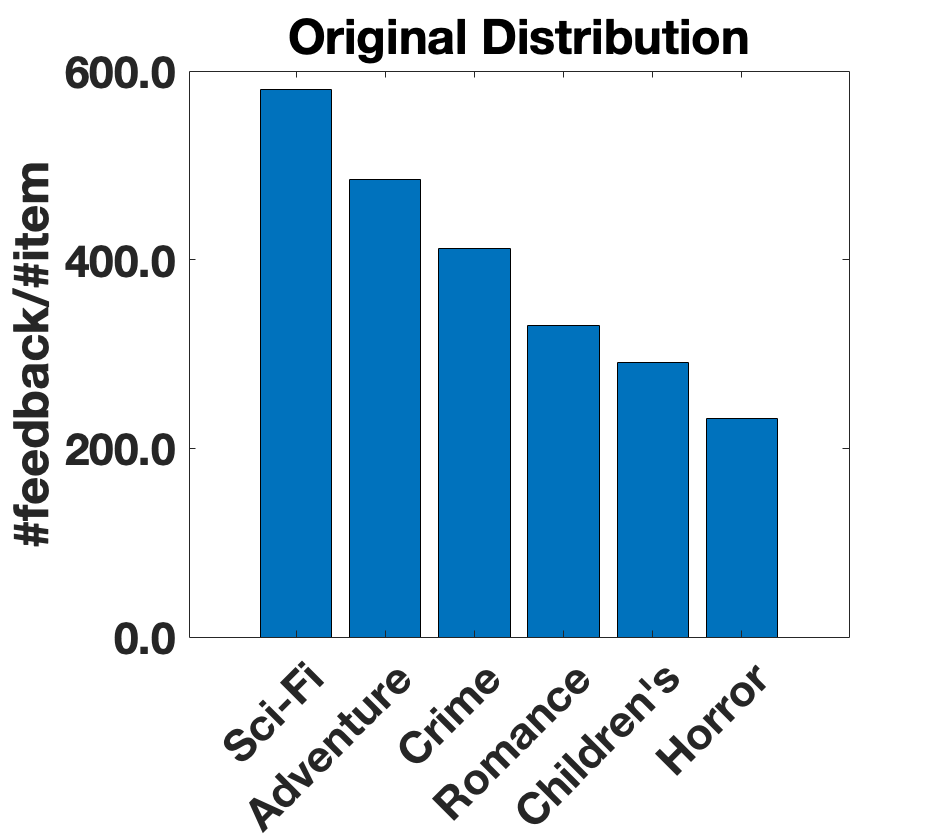}
        \vspace{-16pt}
        \caption{Original.}
        \label{fig:Prelim_Original}
    \end{subfigure}%
    ~ 
    \begin{subfigure}[t]{0.4\textwidth}
        \centering
        \includegraphics[width=1\textwidth]{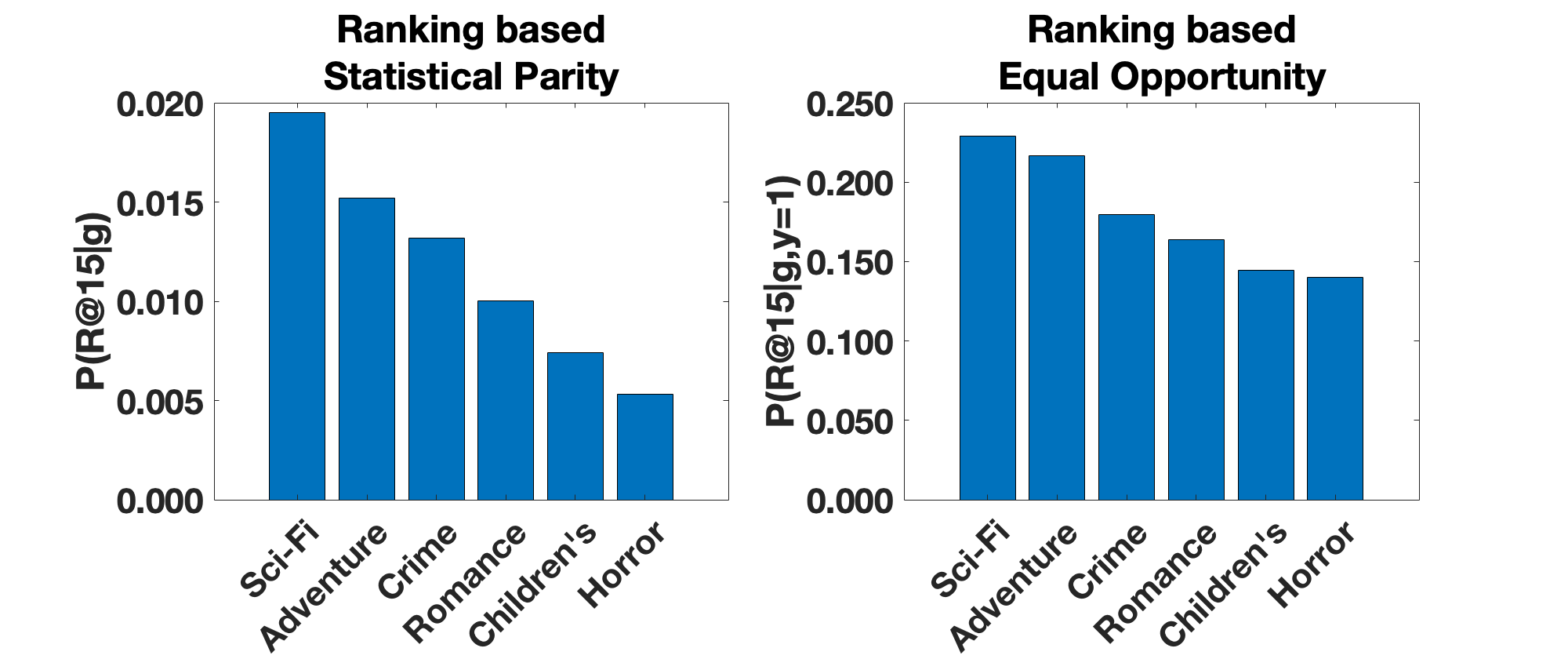}
        \vspace{-16pt}
        \caption{BPR results.}
        \label{fig:Prelim_BPR}
    \end{subfigure}
        ~ 
    \begin{subfigure}[t]{0.4\textwidth}
        \centering
        \includegraphics[width=1\textwidth]{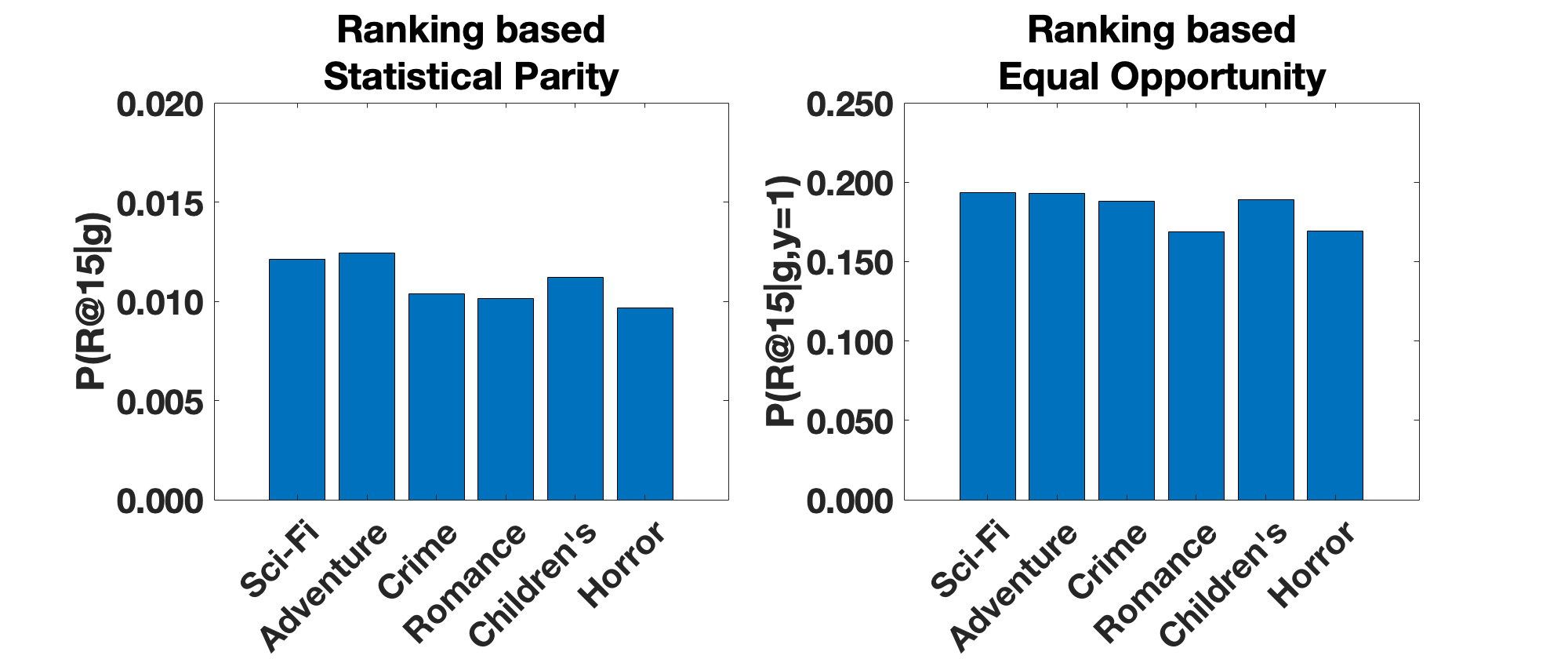}
        \vspace{-16pt}
        \caption{DPR results.}
        \label{fig:Prelim_DPR}
    \end{subfigure}
    \vspace{-12pt}
    \caption{The original distribution of \#feedback/\#item over different groups of ML1M data, and the ranking top15 probability distributions (both statistical parity and equal opportunity based) produced by BPR and proposed DPR.}
    \label{fig:Prelim} 
    \vspace{-10pt}
\end{figure*}

In this section, we empirically show that BPR is vulnerable to imbalanced data and tends to produce unfair recommendation based on metrics RSP and REO. Since there is no standard public dataset related to recommendation fairness with sensitive attributes, we adopt three public real-world datasets that have been extensively used in previous works~\cite{yao2017beyond,he2018adversarial,he2017translation}. However, conclusions we draw should still hold if we analyze the fairness on datasets with sensitive features because the fundamental problem definition and the mechanism leading to unfairness are exactly the same as the experiments in this paper. 

\smallskip
\noindent\textbf{MovieLens 1M (ML1M)}~\cite{harper2016movielens} is a movie rating dataset, where we treat all ratings as positive feedback indicating users are interested in rated movies. We consider the recommendation bias for movie genres of `Sci-Fi', `Adventure', `Crime', `Romance', `Childrens', and `Horror', and remove other films, resulting in $6,036$ users, $1,481$ items, and $526,490$ interactions.

\smallskip
\noindent\textbf{Yelp} (\url{https://www.yelp.com/dataset/challenge}) is a review dataset for businesses. We regard the reviews as the positive feedback showing user interests and only consider restaurant businesses. We investigate the recommendation bias among food genres of `American(New)', `Japanese', `Italian', and `Chinese', resulting in $8,263$ users, $4,420$ items, and $211,721$ interactions.

\smallskip
\noindent\textbf{Amazon}~\cite{mcauley2015image} contains product reviews on the Amazon e-commerce platform. We regard user purchase behaviors as the positive feedback, and consider recommendation bias among product categories of `Grocery', `Office', `Pet', and `Tool', resulting in $4,011$ users, $2,765$ items, and $118,667$ interactions.

\smallskip
Moreover, Table~\ref{table:Datasets_groups} lists the details of each group in the datasets, including the number of items, the number of feedback, and the ratio between them $\frac{\#feedback}{\#item}$. We use this ratio to identify the intrinsic data imbalance. The higher the ratio is, the more this group is favoured by users, and the relative standard deviation of ratios for all groups can indicate overall bias in the dataset. Hence, the Amazon and ML1M datasets contain relatively high bias; and Yelp has lower bias, but American(New) restaurants still have $\frac{\#feedback}{\#item}$ around 1.5 times higher than that of Chinese restaurants.

We run BPR on these datasets and analyze the ranking probability distributions. The detailed model hyper-parameter settings and data splitting are described in Section~\ref{sec:setup}. Table~\ref{table:BPR_Vulnerable} presents $P(R@k|g)$ and $P(R@k|g,y=1)$ for different groups on three datasets by BPR, where we consider $k=5,10,$ and $15$. We also list the metrics $RSP@k$ and $REO@k$. From the table, we have three major observations:

(i) For all datasets, the ranking probabilities are very different among groups, e.g., in ML1M, $P(R@5|g=\text{Sci-Fi})$ is four times higher than $P(R@5|g=\text{Horror})$, and $P(R@5|g=\text{Sci-Fi},y=1)$ is two times higher than $P(R@5|g=\text{Horror},y=1)$. And the high values of $RSP@k$ and $REO@k$ for all $k$ and datasets demonstrate the biased recommendations by BPR.

(ii) The distributions of $P(R@k|g)$ and $P(R@k|g,y=1)$ for all datasets basically follow the distributions of $\frac{\#feedback}{\#item}$ shown in Table~\ref{table:Datasets_groups}, and sometimes the deviations of the ranking probability distributions are even larger than $\frac{\#feedback}{\#item}$ distributions, for example, the relative standard deviation of $P(R@15|g)$ in ML1M is $0.4058$ while that of $\frac{\#feedback}{\#item}$ is $0.3344$, which indicates that BPR preserves or even amplifies the inherent data imbalance.

(iii) As $k$ decreases, the values of $RSP@k$ and $REO@k$ increase. In other words, the results are unfairer for items ranked at top positions. This phenomenon is harmful for recommenders since attention received by items increases rapidly with rankings getting higher \cite{lorigo2008eye}, and top-ranked items get most of attention from users.

Moreover, we also plot the original $\frac{\#feedback}{\#item}$ distribution of ML1M in Figure~\ref{fig:Prelim_Original} and the ranking probability distributions by BPR in Figure~\ref{fig:Prelim_BPR}, which visually confirms our conclusion that BPR inherits data bias and produces unfair recommendations. This conclusion motivates the design of a fairness-aware personalized ranking framework as the models proposed in this paper. Figure~\ref{fig:Prelim_DPR} shows the ranking probability distributions generated by the proposed Debiased Personalized Ranking models, illustrating more evenly distributed and fairer recommendations compared to BPR.

\section{Proposed Method}

\begin{figure}[t!]
% \vspace{-5pt}
\centering
\includegraphics[ width=.9\linewidth ]{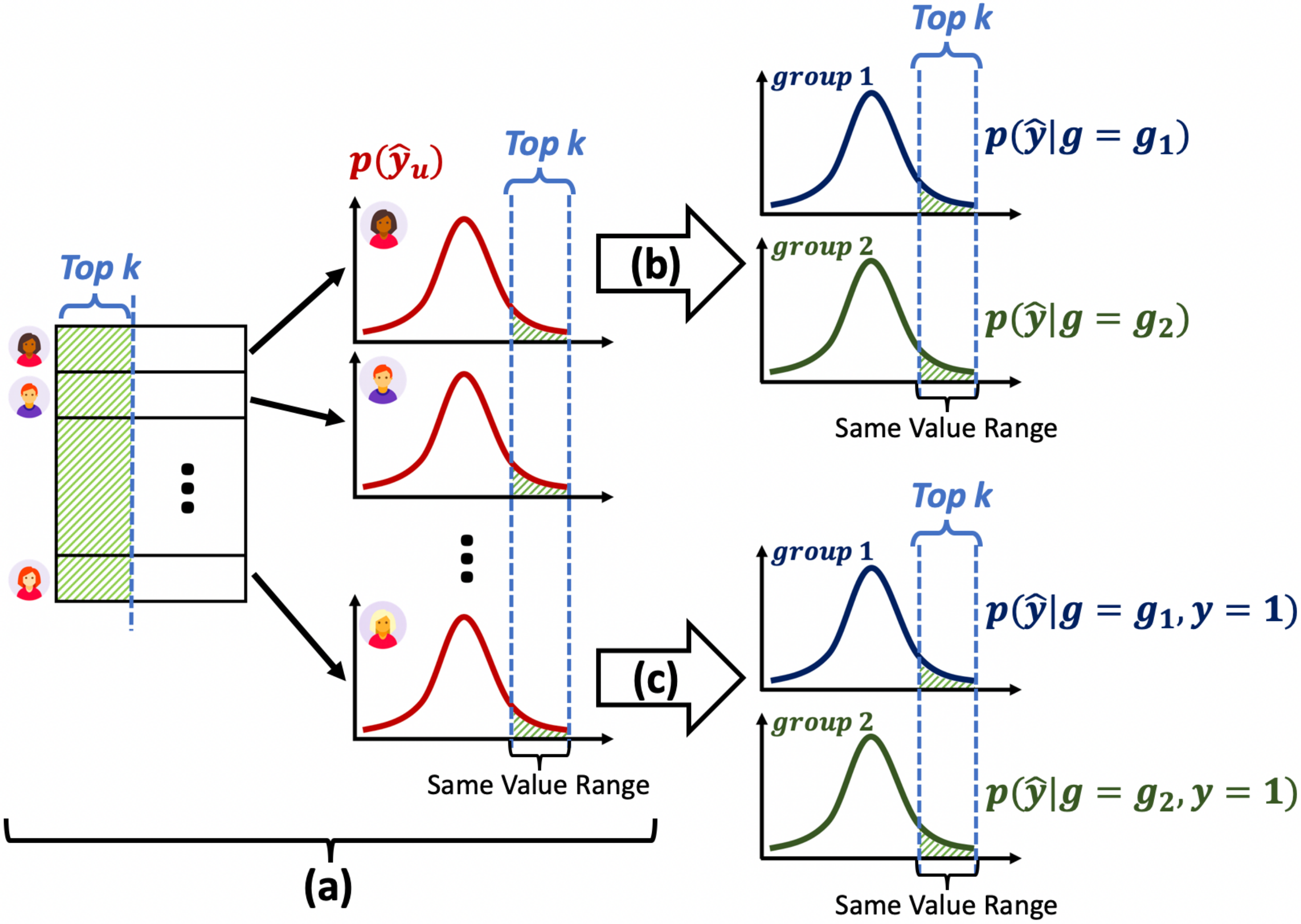} 
\vspace{-10pt} 
\caption{Illustration of the intuition of the proposed DPR.}
% : (a) the value ranges of the scores ranked in the top $k$ are the same for all users if they have the same score distribution; (b) and (c) enforce the score distributions (or conditioned score distribution) to be the same, resulting in fair ranking based on RSP or REO
\label{fig:intuition} 
\vspace{-14pt} 
\end{figure}

Previous works on fairness-aware recommendation~\cite{zhu2018fairness,kamishima2017considerations,kamishima2016model} mainly focus on forcing different groups to have similar score distributions, which cannot necessarily give rise to fair personalized rankings, as shown in Figure~\ref{fig:example}a. One key reason is that users have different rating scales, which means the high score from one user to an item does not necessarily result in a high ranking, and a low score does not lead to a low ranking. Conversely, if every user has an identical score distribution, the value ranges of the scores within the top-$k$ for all users will be the same, as demonstrated in Figure~\ref{fig:intuition}a. Then, the top-$k$ scores in different group score distributions (noted as $p(\widehat{y}|g)$) will also cover the same value range. Last, as illustrated in Figure~\ref{fig:intuition}b, if we enforce identical score distribution for different groups, the proportions of top-$k$ scores in the whole distribution for different groups will be the same, i.e., we have the same probability $p(R@k|g)$ for different groups (the definition of RSP). Similarly, if the positive user-item pairs in different groups have the same score distribution (noted as $p(\widehat{y}|g,y=1)$), we will have the same probability $p(R@k|g,y=1)$ for different groups (the definition of REO), as presented in Figure~\ref{fig:intuition}c. Based on this intuition, the proposed DPR first enhances the score distribution similarity between different groups by adversarial learning, and then normalizes user score distributions to the standard normal distribution by a Kullback-Leibler Divergence (KL) loss. We introduce the two components of DPR and the model training process in the following subsections.

\subsection{Enhancing Score Distribution Similarity}
\label{sec:adversary}

Adversarial learning has been widely applied in supervised learning~\cite{zhang2018mitigating,louppe2017learning,beutel2017data} to enhance model fairness, with theoretical guarantees and state-of-the-art empirical performance. Inspired by these works, we propose to leverage adversarial learning to enhance the score distribution similarity between different groups. We first take the metric RSP as the example to elaborate the proposed method, and then generalize it to REO. Last, we show the advantages of the proposed adversarial learning over previous methods.

\begin{figure}[t!]
% \vspace{-14pt}
\centering
\includegraphics[ width=0.8\linewidth ]{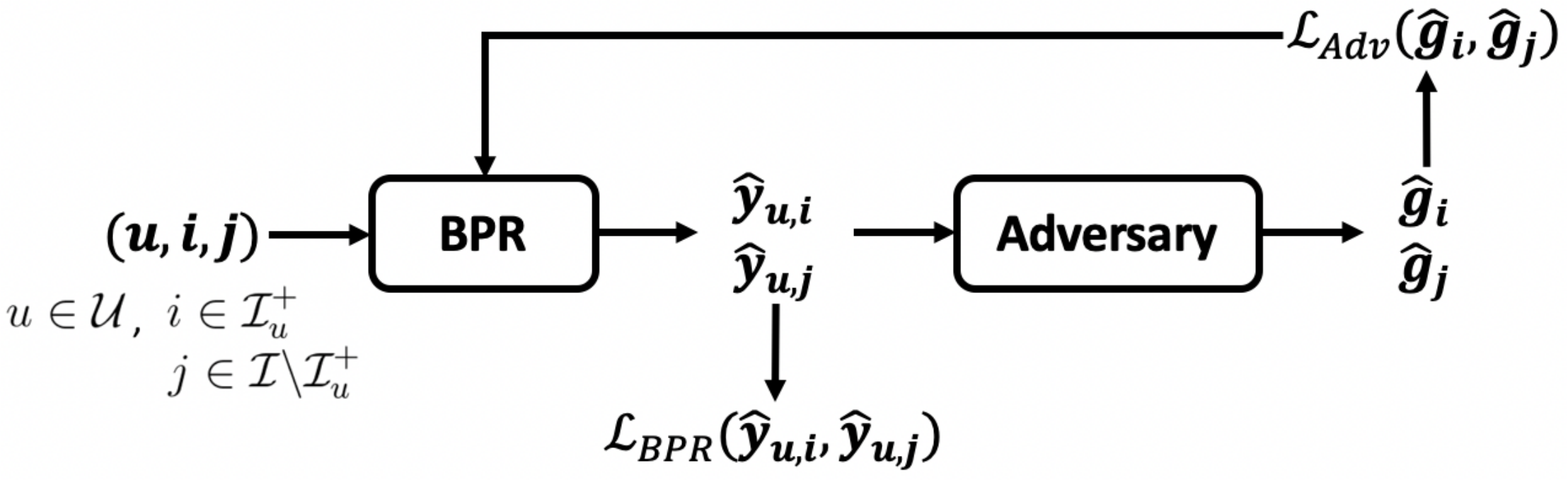} 
\vspace{-10pt} 
\caption{The architecture of the adversarial learning.}
\label{fig:architecture} 
\vspace{-0.5cm} 
\end{figure}

\smallskip
\noindent\textbf{Adversary for RSP.} The intuition of adversarial learning is to play a minimax game between the BPR model and a discriminator. The discriminator is to classify the groups of the items based on the predicted user-item scores by BPR. And BPR does not only need to minimize the recommendation error, but also needs to prevent the discriminator from correctly classifying the groups. If the discriminator cannot accurately recognize the groups given the outputs from BPR, then the predicted score distributions will be identical for different groups. Figure~\ref{fig:architecture} presents the architecture of the adversarial learning framework, where each training user-item pair $(u,i)$ is first input to a conventional BPR model; then the output of BPR, $\widehat{y}_{u,i}$ is fed into a multi-layer perceptron (MLP) to classify the groups $\widehat{\mathbf{g}}_i$ of the given item $i$. $\widehat{\mathbf{g}}_i\in[0,1]^{A}$ is the output of the last layer of MLP activated by the \textit{sigmoid} function, representing the probability of $i$ belonging to each group, e.g., $\widehat{\mathbf{g}}_{i,a}$ means the predicted probability of $i$ belonging to group $g_a$. The MLP is the adversary, which is trained by maximizing the likelihood $\mathscr{L}_{Adv}(\mathbf{g}_i,\widehat{\mathbf{g}}_i)$, and BPR is trained by minimizing the ranking loss shown in Equation~\ref{equ:BPR} as well as minimizing the adversary objective $\mathscr{L}_{Adv}(\mathbf{g}_i,\widehat{\mathbf{g}}_i)$. $\mathbf{g}_i\in\{0,1\}^{A}$ is the ground-truth groups of item $i$, if $i$ is in group $g_a$, $\mathbf{g}_{i,a}=1$, otherwise $0$. We adopt the log-likelihood as the objective function for the adversary:
\begin{equation*}
\begin{aligned}
\centering
\underset{\mathbf{\Psi}}{\text{max}}\,\,
 \mathscr{L}_{Adv}(i)=\sum_{a=1}^{A}(\mathbf{g}_{i,a}log\,\widehat{\mathbf{g}}_{i,a}+(1-\mathbf{g}_{i,a})log\,(1-\widehat{\mathbf{g}}_{i,a})),
\end{aligned}
\label{equ:Adv}
\end{equation*}
where we denote $\mathscr{L}_{Adv}(\mathbf{g}_i,\widehat{\mathbf{g}}_i)$ as $\mathscr{L}_{Adv}(i)$ for short, and $\mathbf{\Psi}$ is the parameters of the MLP adversary. Combined with the BPR model, the objective function can be formulated as:
\begin{equation}
\begin{aligned}
\centering
\underset{\mathbf{\Theta}}{\text{min}}\,\underset{\mathbf{\Psi}}{\text{max}}\,\,&\sum_{u\in\mathcal{U}}\sum_{\substack{i\in\mathcal{I}_{u}^+\\j\in\mathcal{I}{\setminus}\mathcal{I}_u^+}} \mathscr{L}_{BPR}(u,i,j)+\alpha(\mathscr{L}_{Adv}(i)+\mathscr{L}_{Adv}(j)),\\
\text{where}\,\,&\mathscr{L}_{BPR}(u,i,j)=-ln\, \sigma(\widehat{y}_{u,i}-\widehat{y}_{u,j}) + \dfrac{\lambda_{\mathbf{\Theta}}}{2}\lVert\mathbf{\Theta}\rVert^2_{\text{F}},
\end{aligned}
\label{equ:BPR_Adv_SP}
\end{equation}
and $\alpha$ is the trade-off parameter to control the strength of the adversarial component.

\smallskip
\noindent\textbf{Adversary for REO.}
As for REO, we demand the score distributions of positive user-item pairs rather than all the user-item pairs to be identical for different groups. Therefore, instead of feeding both scores for positive and sampled negative user-item pairs $\widehat{y}_{u,i}$ and $\widehat{y}_{u,j}$, we only need to feed $\widehat{y}_{u,i}$ into the adversary as:
\begin{equation}
\begin{aligned}
\centering
\underset{\mathbf{\Theta}}{\text{min}}\,\underset{\mathbf{\Psi}}{\text{max}}\,\,& \sum_{u\in\mathcal{U}}\sum_{\substack{i\in\mathcal{I}_{u}^+\\j\in\mathcal{I}{\setminus}\mathcal{I}_u^+}} \mathscr{L}_{BPR}(u,i,j)+\alpha\mathscr{L}_{Adv}(i).
\end{aligned}
\label{equ:BPR_Adv_EO}
\end{equation}

\smallskip
\noindent\textbf{Advantages of adversarial learning.}
There are two existing approaches to achieve a similar effect: a regularization-based method~\cite{kamishima2013efficiency,kamishima2017considerations,yao2017beyond}; and a latent factor manipulation method~\cite{zhu2018fairness}. The advantages of the proposed adversarial learning over previous works can be summarized as: (i) it can provide more effective empirical performance than other methods, which will be further demonstrated in Section~\ref{sec:RQ2}; (ii) it is flexible to swap in different bias metrics (beyond just RSP and REO); (iii) it can handle multi-group circumstances; and (iv) it is not coupled with any specific recommendation models and can be easily adapted to methods other than BPR (such as more advanced neural networks).

\subsection{Individual User Score Normalization}
\label{sec:KL-loss}
After the enforcement of distribution similarity, the next step towards personalized ranking fairness is to normalize the score distribution for each user. We can assume the score distribution of every user follows the normal distribution because based on the original BPR paper~\cite{rendle2009bpr}, every factor in the user or item latent factor vector follows a normal distribution, then $\mathbf{P}_u^\top\mathbf{Q}_i$ (for a given user $u$, $\mathbf{P}_u$ is a constant and $\mathbf{Q}_i$ is a vector of normal random variables) follows a normal distribution as well. Thus we can normalize the score distribution of each user to the standard normal distribution by minimizing the KL Divergence between the score distribution of each user and a standard normal distribution as the KL-loss:
\begin{equation*}
\begin{aligned}
\centering
 \mathscr{L}_{KL}=\sum_{u\in\mathcal{U}}D_{\text{KL}}(q_\mathbf{\Theta}(u)||\mathcal{N}(0,1)),
\end{aligned}
\label{equ:KL-loss}
\end{equation*}
where $q_\mathbf{\Theta}(u)$ is the empirical distribution of predicted scores for user $u$, and $D_{\text{KL}}(\cdot||\cdot)$ computes KL Divergence between two distributions.

\subsection{Model Training}

Combining the KL-loss with Equation~\ref{equ:BPR_Adv_SP} leads to the complete DPR models to optimize RSP, noted as DPR-RSP:
\begin{equation*}
\begin{aligned}
\centering
\underset{\mathbf{\Theta}}{\text{min}}\,\underset{\mathbf{\Psi}}{\text{max}}\sum_{u\in\mathcal{U}}\!\!\sum_{\substack{i\in\mathcal{I}_{u}^+\\j\in\mathcal{I}{\setminus}\mathcal{I}_u^+}}\!\! (\mathscr{L}_{BPR}(u,i,j)+\alpha(\mathscr{L}_{Adv}(i)+\mathscr{L}_{Adv}(j)))+\beta\mathscr{L}_{KL},
\end{aligned}
\label{equ:DPR_SP}
\end{equation*}
where $\beta$ is the trade-off parameter to control the strength of KL-loss. Similarly, we can optimize REO by combining KL-loss with Equation~\ref{equ:BPR_Adv_EO} to arrive at a DPR-REO model as well. Note that although the proposed DPR is built with BPR as the model foundation, it is in fact flexible enough to be adapted to other recommendation algorithms, such as more advanced neural networks~\cite{he2017neural}.

Then, we train the model in a mini-batch manner. Generally, during model training, there are two phases in each epoch: first we update weights in the MLP adversary to maximize the classification objective, then update BPR to minimize the pairwise ranking loss, classification objective and KL-loss all together. Concretely, following the adversarial training process proposed in \cite{louppe2017learning}, in each epoch, we first update the MLP adversary by the whole dataset (in a stochastic way), then update BPR by one mini-batch, which empirically leads to fast convergence. And in practice, we usually first pre-train the BPR model for several epochs and then add in the adversarial training part. The detailed training process including pseudo code is introduced in Appendix~\ref{sec:model_training}.

\section{Experiments}
\label{sec:experiment}

In this section, we empirically evaluate the proposed model w.r.t. the two proposed bias metrics as well as the recommendation quality. We aim to answer three key research questions: \textbf{RQ1} What are the effects of the proposed KL-loss, adversary, and the complete model DPR on recommendations? \textbf{RQ2} How does the proposed DPR perform compared with other state-of-the-art debiased models from the perspectives of improving fairness and recommendation quality preserving? and \textbf{RQ3} How do hyper-parameters affect the DPR framework? 

\subsection{Datasets}
\label{sec:datasets}

\begin{table}[t!]\small
  \begin{center}
  \renewcommand{\arraystretch}{.9}
    \begin{tabular}[0.1\textwidth]{ c|cccc}  
\hline\hline
                        		& \#Users   & \#Items   & \#Ratings   & Density  \\  \hline 
ML1M-2     		                & 5,562     & 543       & 215,549     & 7.14\%          \\  
Yelp-2            	    		& 6,310     & 2,834     & 117,978     & 0.66\%         \\
Amazon-2   	                    & 3,845     & 2,487     & 84,656     & 0.89\%         \\   
\hline\hline
    \end{tabular}
  \end{center}
%   \vspace{.2cm}
        \caption{Characteristics of the three 2-group datasets.}
        \label{table:Datasets-2} 
\vspace{-20pt}
\end{table}

The three datasets used in the experiments have been introduced in Section~\ref{sec:BPR_Experiment}. Since the state-of-the-art baselines can only work for binary group cases, to answer \textbf{RQ2}, we create subsets keeping the most popular and least popular groups in the original datasets: \textbf{ML1M-2} (`Sci-Fi' vs `Horror'), \textbf{Yelp-2} (`American(New)' vs. `Chinese'), and \textbf{Amazon-2} (`Grocery' vs. `Tool'). The specifics of the 2-group datasets are presented in Table~\ref{table:Datasets-2}. All datasets are randomly split into 60\%, 20\%, 20\% for training, validation, and test sets. Note that there is no standard public dataset with sensitive features, thus we use public datasets for general recommendation scenarios to evaluate the performance of enhancing RSP-based fairness. However, conclusions we draw should still hold if we analyze the fairness-enhancement performance on datasets with sensitive features because the fundamental problem definition and the mechanism leading to unfairness are exactly the same as the experiments in this paper.

\subsection{Experimental Setup}
\label{sec:setup}

\noindent\textbf{Metrics.}
In the experiments, we need to consider both recommendation quality and recommendation fairness. For the recommendation fairness, we report $RSP@k$ and $REO@k$ as described in Section~\ref{sec:metrics}. As for the recommendation quality we adopt $F1@k$. We report the results with $k=5,10,$ and $15$. Note that we also measure NDCG in the experiments, which shows the same pattern as F1, hence we only report $F1@k$ for conciseness.

\medskip
\noindent\textbf{Baselines.}
\label{sec:baselines}
We compare the proposed DPR with unfair method BPR shown in Section~\ref{sec:BPR} and two state-of-the-art fairness-aware recommendation methods:

\textbf{FATR}~\cite{zhu2018fairness}. This is a tensor-based method, which enhances the score distribution similarity for different groups by manipulating the latent factor matrices. We adopt the 2D matrix version of this approach. Note that FATR is designed for statistical parity based metric, hence we do not have high expectation for the performance w.r.t. equal opportunity. 

\textbf{Reg}~\cite{kamishima2013efficiency,kamishima2017considerations,yao2017beyond}. The most commonly used fairness-aware method for two-group scenarios, which penalizes recommendation difference by minimizing a regularization term. Following \cite{kamishima2017considerations}, we adopt the squared difference between the average scores of two groups for all items as the regularization to improve RSP, denoted as \textbf{Reg-RSP}. For REO, we adopt the squared difference between the average scores of positive user-item pairs as the regularization, denoted as \textbf{Reg-REO} (it is similar to DPR-REO but enhances the distribution similarity by static regularization rather than adversary).

To have a fair comparison, we modify the loss functions of all baselines to the BPR loss in Equation~\ref{equ:BPR}. Moreover, to align the baselines with the bias metrics for ranking, we further add the proposed KL-loss introduced in Section~\ref{sec:KL-loss} to both baselines. The new objective functions of the baselines are introduced in detail in Appendix~\ref{sec:baseline_objective_function}.

% The new objective functions of the baselines are introduced in detail in Appendix~\ref{sec:baseline_objective_function}.

\medskip
\noindent\textbf{Reproducibility.} Code and data for this work can be found at \url{https://github.com/Zziwei/Item-Underrecommendation-Bias}. We implement the proposed model using Tensorflow~\cite{abadi2016tensorflow} and adopt Adam~\cite{kingma2014adam} optimization algorithm. We tune the hyper-parameters of the models involved by the validation set, the basic rules are: (i) we search the hidden dimension over $\{10, 20, 30, 40, 50, 60, 70, 80\}$; (ii) search the $L_2$ regularizer $\lambda_\mathbf{\Theta}$ over $\{0.01, 0.05, 0.1, 0.5, 1.0\}$; (iii) search the adversary regularizer $\alpha$ over range $[500, 10000]$ with step 500; (iv) search the KL-loss regularizer $\beta$ over range $[10, 70]$ with step 10; and (v) search the model specific weight in FATR over $\{0.01, 0.05, 0.1, 0.5, 1.0\}$, and model specific weight for Reg-RSP and Reg-REO over the range $[1000, 10000]$ with step 2000. Note that selections of $\alpha$ and $\beta$ should consider the balance between recommendation quality and recommendation fairness.

There are two sets of experiments: experiments over multi-group datasets (ML1M, Yelp, and Amazon) to answer \textbf{RQ1} and \textbf{RQ3}; and experiments over binary-group datasets (ML1M-2, Yelp-2, and Amazon-2) to answer \textbf{RQ2}.

In the first set of experiments, for all three datasets: we set 20 as the hidden dimensions for BPR, DPR-RSP, and DPR-REO; we set the learning rate 0.01 for BPR, and $\eta_{BPR}$ 0.01 as well for DPR-RSP and DPR-REO. For all methods, we set $\lambda_\mathbf{\Theta}=0.1$ for ML1M and Amazon; set $\lambda_\mathbf{\Theta}=0.05$ for Yelp. As for adversary learning rate $\eta_{Adv}$, we set 0.005 for ML1M and Yelp, 0.001 for Amazon. For all three datasets, we set $\alpha=5000$ for DPR-RSP. As for DPR-REO, we set $\alpha=1000$ for ML1M, 5000 for Yelp, and 10000 for Amazon.

In the second set of experiments, we set different hidden dimensions for different datasets, but for the same dataset all methods have the same dimension: we set 10 for ML1M-2, 40 for Yelp-2, and 60 for Amazon-2. We set the learning rate 0.01 for baselines, and 0.01 as $\eta_{BPR}$ for DPR-RSP and DPR-REO. As for adversary learning rate $\eta_{Adv}$, we set 0.005 for all three datasets.

For all methods in all experiments, we have negative sampling rate 5 and mini-batch size 1024. For all fairness-aware methods, we set $\beta=30$. And we adopt a 4-layer MLP with 50 neurons with ReLU activation function in each layer as the adversary for DPR.

\subsection{RQ1: Effects of Model Components}
\label{sec:RQ1}
In this subsection, we aim to answer three questions: whether the KL-loss can effectively normalize user score distribution? whether the adversary can effectively enhance score distribution similarity among groups? and whether  DPR-RSP and DPR-REO can effectively improve the fairness metrics RSP and REO?

\begin{table}[t!]\small
  \begin{center}
\renewcommand{\arraystretch}{.9}
    \begin{tabular}[0.1\textwidth]{ c|cccc}  
    \hline\hline
                & ML1M    & Yelp    & Amazon  \\ \hline 
BPR     		& 0.1540  & 0.0808  & 0.0836  \\  
BPR w/ KL-loss  & 0.0571  & 0.0254  & 0.0313  \\ \hline 
$\Delta$        & \textbf{-62.92\%} & \textbf{-68.56\%} & \textbf{-62.56\%}   \\ 
\hline\hline
    \end{tabular}
  \end{center}
%   \vspace{.2cm}
        \caption{Comparison between BPR w/o KL-loss for JS Divergences among user score distributions over three datasets.}
        \label{table:RQ1} 
\vspace{-15pt}
\end{table}

\begin{figure}[t!]
    \centering
    \begin{subfigure}[t]{0.22\textwidth}
        \centering
        \includegraphics[width=1\textwidth]{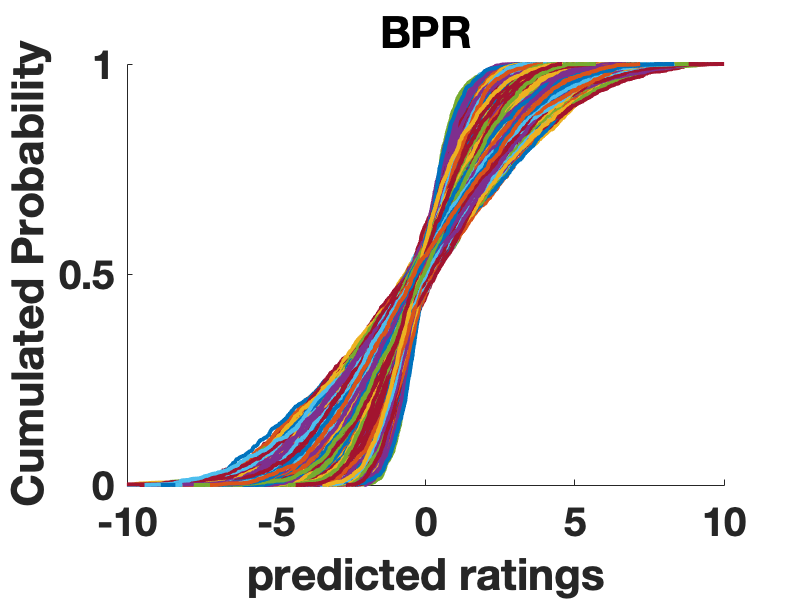}
        %\vspace{-7pt}
        \label{fig:RQ1_BPR_KL}
    \end{subfigure}%
        ~ %\hspace*{15pt}
    \begin{subfigure}[t]{0.22\textwidth}
        \centering
        \includegraphics[width=1\textwidth]{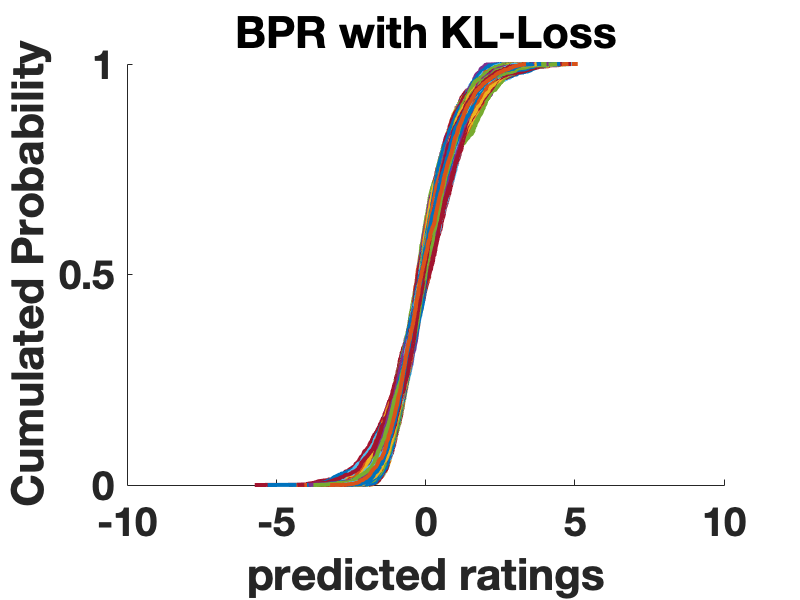}
        %\vspace{-7pt}
        \label{fig:RQ1_KL_KL}
    \end{subfigure}
    \vspace{-18pt}
    \caption{CDFs of user score distributions predicted by BPR and BPR with KL-loss over ML1M dataset.}
    \label{fig:RQ1_KL} 
    \vspace{-5pt}
\end{figure}

\smallskip
\noindent\textbf{Effects of KL-loss.} The KL-loss is to normalize the user score distribution. Hence, we adopt the Jensen-Shannon Divergence (JS Divergence) to measure the deviation between user score distributions, where lower JS Divergence indicates that the user score distributions are normalized better. We compare BPR and BPR with KL-loss over all three datasets, the results are shown in Table \ref{table:RQ1}, and the improvement rates (noted as $\Delta$) are also calculated. We can observe that with the KL-loss, the divergence among user score distributions is largely reduced, demonstrating the effectiveness of KL-loss. To better show the effects of KL-loss, we visualize the score distribution for every user produced by BPR with and without KL-loss for ML1M in Figure \ref{fig:RQ1_KL}, where each curve represents the Cumulative Distribution Function (CDF) of a single user's scores. The closely centralized CDFs in the right figure verify the effectiveness of the proposed KL-loss.

\begin{figure}[t!]
    \centering
    \begin{subfigure}[t]{0.22\textwidth}
        \centering
        \includegraphics[width=1\textwidth]{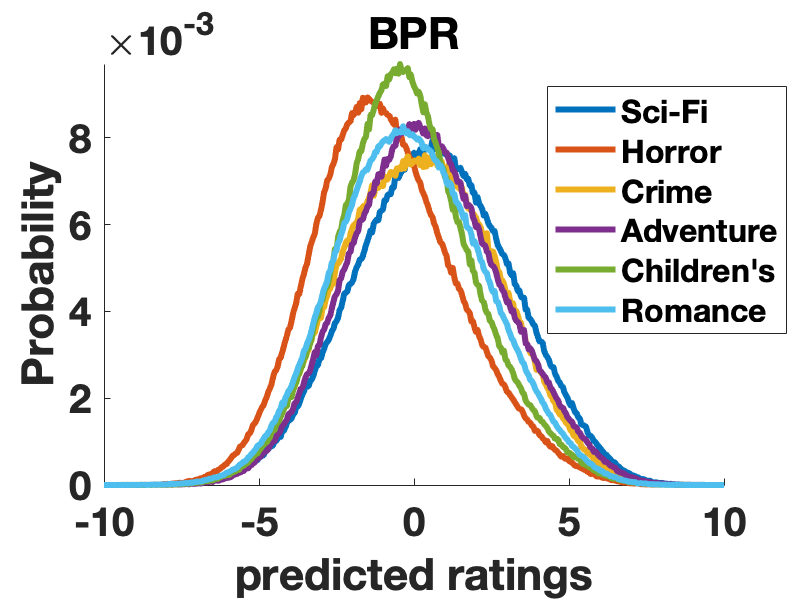}
        %\vspace{-7pt}
        \label{fig:RQ1_adv_SP_BPR}
    \end{subfigure}%
        ~ %\hspace*{15pt}
    \begin{subfigure}[t]{0.22\textwidth}
        \centering
        \includegraphics[width=1\textwidth]{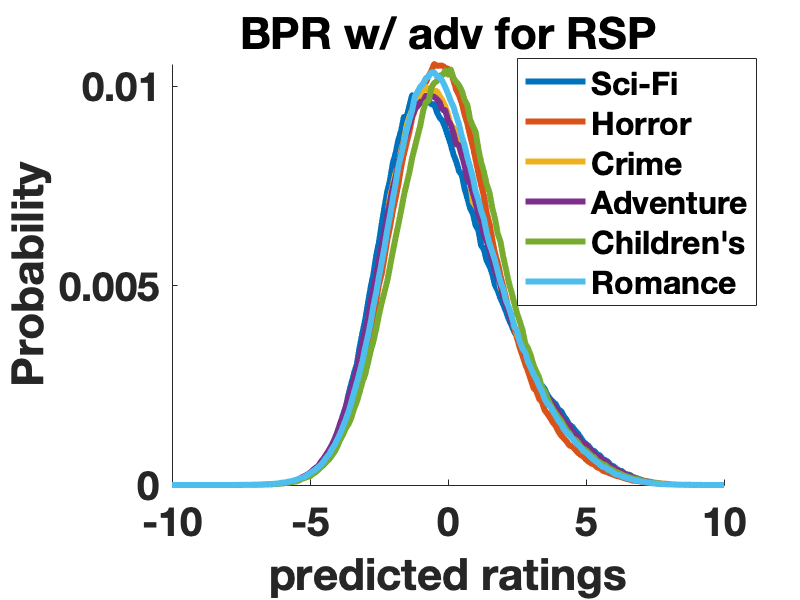}
        %\vspace{-7pt}
        \label{fig:RQ1_adv_SP_FPR}
    \end{subfigure}
    \vspace{-18pt}
    \caption{PDFs of $p(\widehat{y}|g)$ for different groups by BPR and BPR w/ adv for RSP over ML1M dataset.}
    \label{fig:RQ1_adv_SP} 
    \vspace{-5pt}
\end{figure}

\begin{table}[t!]\small
  \begin{center}
\renewcommand{\arraystretch}{.9}
\begin{tabular}{cc|c|c|c}
\hline\hline
                                         &            & ML1M     & Yelp     & Amazon   \\ \hline
\multicolumn{1}{c|}{\multirow{3}{*}{RSP setting}} & BPR        & 0.0222   & 0.0011   & 0.0215   \\
\multicolumn{1}{c|}{}                    & BPR w/ adv & 0.0090   & 0.0004   & 0.0046   \\ \cline{2-5} 
\multicolumn{1}{c|}{}                    & $\Delta$   & \textbf{-59.46\%} & \textbf{-63.64\%} & \textbf{-78.60\%} \\ \hline
\multicolumn{1}{c|}{\multirow{3}{*}{REO setting}} & BPR        & 0.0128   & 0.0045   & 0.0378   \\
\multicolumn{1}{c|}{}                    & BPR w/ adv & 0.0047   & 0.0041   & 0.0087   \\ \cline{2-5} 
\multicolumn{1}{c|}{}                    & $\Delta$   & \textbf{-63.28\%} & \textbf{-8.89\%}  & \textbf{-76.98\%} \\ \hline\hline
\end{tabular}
  \end{center}
%   \vspace{-2pt}
        \caption{Comparison between BPR and BPR w/ adv for JS Divergences of score distribution among different groups.}
\label{table:RQ1_adv} 
\vspace{-15pt}
\end{table}

\smallskip
\noindent\textbf{Effects of Adversary.} The adversary in DPR is to enhance the score distribution similarity among different groups. To evaluate the effectiveness of the adversarial learning, we compare the performances of BPR and BPR with adversary for both metrics (noted as \textit{BPR w/ adv for RSP} and \textit{BPR w/ adv for REO}). More specifically, we compare BPR with BPR w/ adv for RSP w.r.t. JS Divergence among $p(\widehat{y}|g)$ for different groups, and compare BPR with BPR w/ adv for REO w.r.t. JS Divergence among $p(\widehat{y}|g,y=1)$ for different groups. Results are shown in Table~\ref{table:RQ1_adv}, where the top three rows are calculated on all user-item pairs not in the training set (fit the RSP setting), the bottom three rows are calculated on user-item pairs only in the test set (fit the REO setting). The table demonstrates the extraordinary effectiveness of the proposed adversarial learning for enhancing distribution similarity under both settings. To further validate this conclusion, we visualize the distributions of $p(\widehat{y}|g)$ for different groups from ML1M in Figure~\ref{fig:RQ1_adv_SP} (distributions of $p(\widehat{y}|g,y=1$ have the same pattern), where the Probability Distribution Function (PDF) of every group's score distribution is plot as a single curve. We can find that PDFs by BPR w/ adv are close to each other, while PDFs by the ordinary BPR differ considerably.

\begin{table}[t!]\small
  \begin{center}
\renewcommand{\arraystretch}{.9}
\begin{tabular}{lcl|c|c|c}
\hline\hline
                                             & \multicolumn{1}{l}{}     &    & ML1M     & Yelp     & Amazon   \\ \hline
\multicolumn{1}{l|}{\multirow{3}{*}{F1@15}}  & \multicolumn{2}{c|}{BPR}      & 0.1520   & 0.0371   & 0.0230   \\
\multicolumn{1}{l|}{}                        & \multicolumn{2}{c|}{DRP-RSP}  & 0.1439   & 0.0354   & 0.0221   \\ \cline{2-6} 
\multicolumn{1}{l|}{}                        & \multicolumn{2}{c|}{$\Delta$} & \textbf{-5.31}\%  & \textbf{-4.32}\%  & \textbf{-3.90}\%  \\ \hline
\multicolumn{1}{l|}{\multirow{3}{*}{RSP@15}} & \multicolumn{2}{c|}{BPR}      & 0.4058   & 0.2522   & 0.4155   \\
\multicolumn{1}{l|}{}                        & \multicolumn{2}{c|}{DPR-RSP}  & 0.0936   & 0.0856   & 0.0607   \\ \cline{2-6} 
\multicolumn{1}{l|}{}                        & \multicolumn{2}{c|}{$\Delta$} & \textbf{-76.92\%} & \textbf{-66.07\%} & \textbf{-85.40\%} \\ \hline\hline
\end{tabular}
  \end{center}
%   \vspace{-2pt}
        \caption{Comparison between BPR and DPR-RSP w.r.t. $F1@15$ and $RSP@15$ over three datasets.}
\label{table:RQ1_FPR_SP} 
\vspace{-22pt}
\end{table}

\begin{table}[t!]\small
  \begin{center}
\renewcommand{\arraystretch}{.9}
\begin{tabular}{lcl|c|c|c}
\hline\hline
                                             & \multicolumn{1}{l}{}     &    & ML1M     & Yelp     & Amazon   \\ \hline
\multicolumn{1}{l|}{\multirow{3}{*}{F1@15}}  & \multicolumn{2}{c|}{BPR}      & 0.1520   & 0.0371   & 0.0230   \\
\multicolumn{1}{l|}{}                        & \multicolumn{2}{c|}{DRP-REO}  & 0.1527   & 0.0363   & 0.0208   \\ \cline{2-6} 
\multicolumn{1}{l|}{}                        & \multicolumn{2}{c|}{$\Delta$} & \textbf{+0.49\%}  & \textbf{-1.94\%}  & \textbf{-9.81\%}  \\ \hline
\multicolumn{1}{l|}{\multirow{3}{*}{REO@15}} & \multicolumn{2}{c|}{BPR}      & 0.1893   & 0.2225   & 0.6217   \\
\multicolumn{1}{l|}{}                        & \multicolumn{2}{c|}{DPR-REO}  & 0.0523   & 0.0874   & 0.3577   \\ \cline{2-6} 
\multicolumn{1}{l|}{}                        & \multicolumn{2}{c|}{$\Delta$} & \textbf{-72.38\%} &\textbf{ -60.73\%} & \textbf{-42.47\%} \\ \hline\hline
\end{tabular}
  \end{center}
%   \vspace{-2pt}
        \caption{Comparison between BPR and DPR-REO w.r.t. $F1@15$ and $REO@15$ over three datasets.}
\label{table:RQ1_FPR_EO} 
\vspace{-15pt}
\end{table}

\smallskip
\noindent\textbf{Effects of DPR.} The effects of the complete DPR should be evaluated from the perspectives of both recommendation quality and recommendation fairness. We first investigate the performance of DPR-RSP. $F1@15$ and $RSP@15$ results of both BPR and DPR-RSP over three datasets are listed in Table~\ref{table:RQ1_FPR_SP}, where the change rates for them are calculated. From the table we have three observations: (i) DPR-RSP improves the fairness over BPR greatly (decreases $RSP@15$ by 76\% on average); (ii) DPR-RSP effectively preserves the recommendation quality (only drops $F1@15$ by 4\% on average); and (iii) for different datasets with different degrees of data imbalance, DPR-RSP can reduce the unfairness to a similar level ($RSP@15$ for three datasets by DPR-RSP are all smaller than $0.1$).

Similar conclusions can be drawn for DPR-REO based on Table~\ref{table:RQ1_FPR_EO}, where comparison between BPR and DPR-REO w.r.t. $F1@15$ and $REO@15$ are listed. We can observe that DPR-REO is able to decrease metric $REO@15$ to a great extent while preserving high $F1@15$ as well. Generally speaking, DPR-REO demands less recommendation quality sacrifice because the definition of REO is less stringent and enhancing fairness is easier to achieve than RSP. However, there is one exception that in Amazon dataset, DPR-REO drops $F1@15$ by $9.8\%$. It may be because for the Amazon dataset, every item group has its own collection of users, and there are few users giving feedback to more than one group, which exerts difficulty for DPR-REO training.

\subsection{RQ2: Comparison with Baselines}
\label{sec:RQ2}
\begin{table}[t!]\small
  \begin{center}
\renewcommand{\arraystretch}{.9}
\begin{tabular}{cc|c|c|c}
\hline\hline
                                         &        & ML1M-2          & Yelp-2          & Amazon-2        \\ \hline
\multicolumn{1}{c|}{\multirow{5}{*}{RSP setting}} & BPR    & 0.0564          & 0.0034          & 0.0514          \\
\multicolumn{1}{c|}{}                    & FATR     & \textbf{0.0218} & 0.0027          & \textbf{0.0332} \\
\multicolumn{1}{c|}{}                    & Reg-RSP & 0.0276          & \textbf{0.0026} & 0.0378          \\
\multicolumn{1}{c|}{}                    & DPR-RSP & \textbf{0.0155} & \textbf{0.0020} & \textbf{0.0079} \\ \cline{2-5} 
\multicolumn{1}{c|}{}                    & $\Delta$  & -28.90\%        & -23.08\%        & -76.20\%        \\ \hline
\multicolumn{1}{c|}{\multirow{5}{*}{REO setting}} & BPR    & 0.0422          & 0.0216          & 0.1531          \\
\multicolumn{1}{c|}{}                    & FATR     & \textbf{0.0044} & 0.0078          & 0.1844          \\
\multicolumn{1}{c|}{}                    & Reg-REO & 0.0179          & \textbf{0.0062} & \textbf{0.0219} \\
\multicolumn{1}{c|}{}                    & DPR-REO & \textbf{0.0011} & \textbf{0.0018} & \textbf{0.0038} \\ \cline{2-5} 
\multicolumn{1}{c|}{}                    & $\Delta$  & -75.00\%        & -70.97\%        & -82.65\%  \\ \hline\hline
\end{tabular}
  \end{center}
%   \vspace{-2pt}
        \caption{Comparison between DPR and baselines for JS Divergences of score distribution among groups.}
\label{table:RQ2} 
\vspace{-25pt}
\end{table}

\begin{figure*}[t!]
    \centering
    \begin{subfigure}[t]{0.155\textwidth}
        \centering
        \includegraphics[width=1\textwidth]{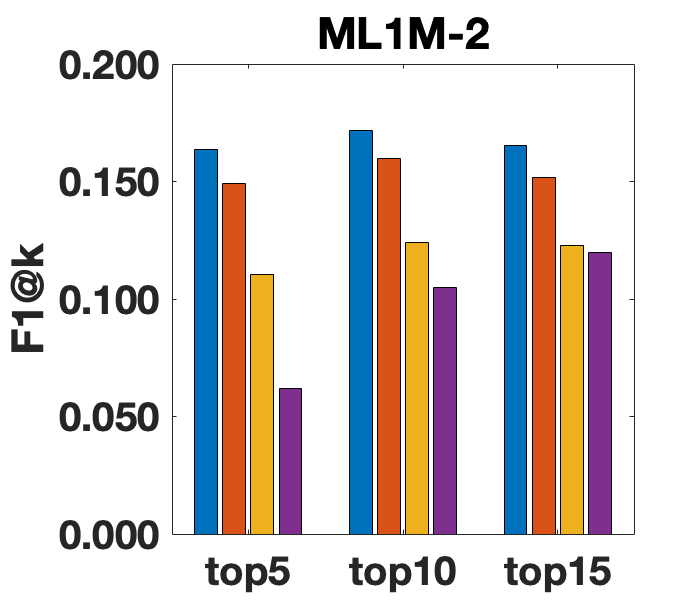}
        %\vspace{-7pt}
        \label{fig:RQ2_SP_F_ML1M}
    \end{subfigure}
        ~ 
    \begin{subfigure}[t]{0.155\textwidth}
        \centering
        \includegraphics[width=1\textwidth]{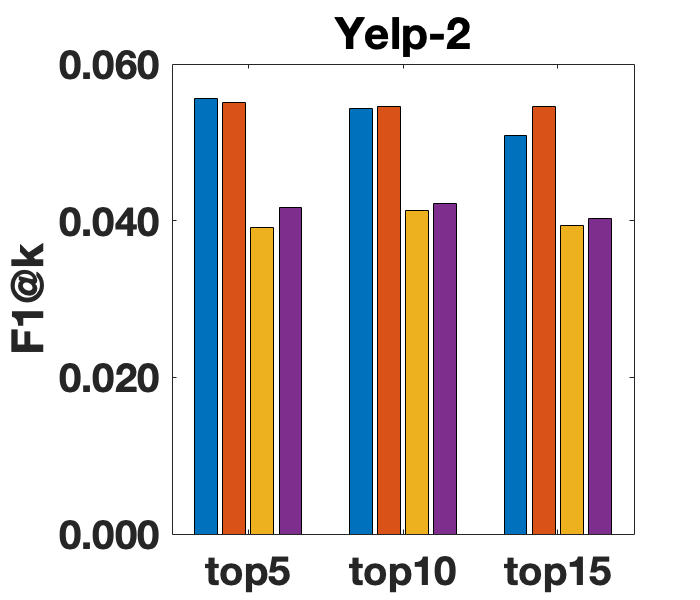}
        %\vspace{-7pt}
        \label{fig:RQ2_SP_F_Yelp}
    \end{subfigure}
            ~ 
    \begin{subfigure}[t]{0.155\textwidth}
        \centering
        \includegraphics[width=1\textwidth]{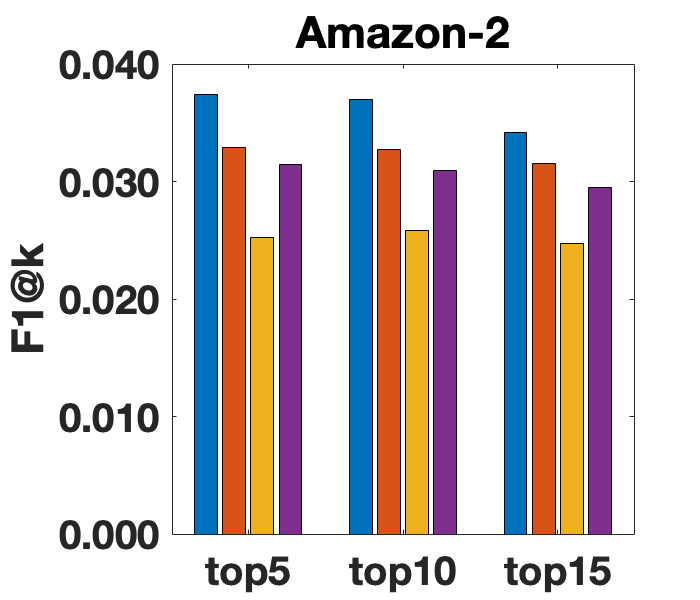}
        %\vspace{-7pt}
        \label{fig:RQ2_SP_F_Amazon}
    \end{subfigure}
            ~ 
    \begin{subfigure}[t]{0.155\textwidth}
        \centering
        \includegraphics[width=1\textwidth]{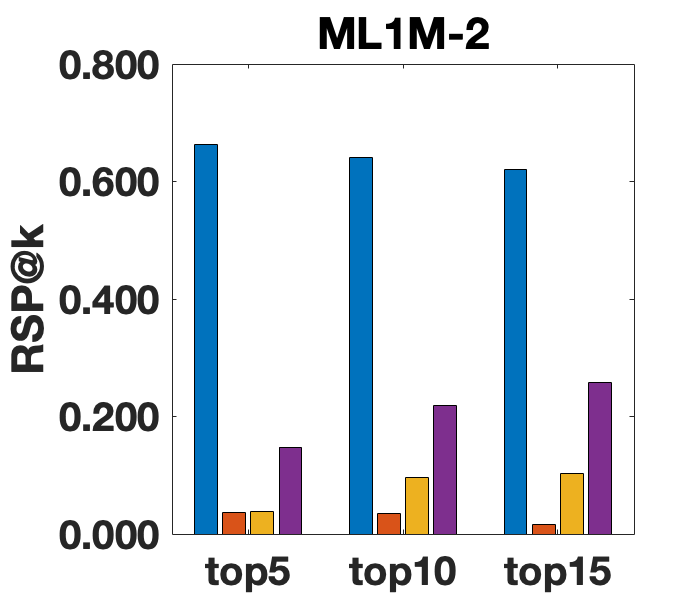}
        %\vspace{-7pt}
        \label{fig:RQ2_SP_SP_ML1M}
    \end{subfigure}
            ~ 
    \begin{subfigure}[t]{0.155\textwidth}
        \centering
        \includegraphics[width=1\textwidth]{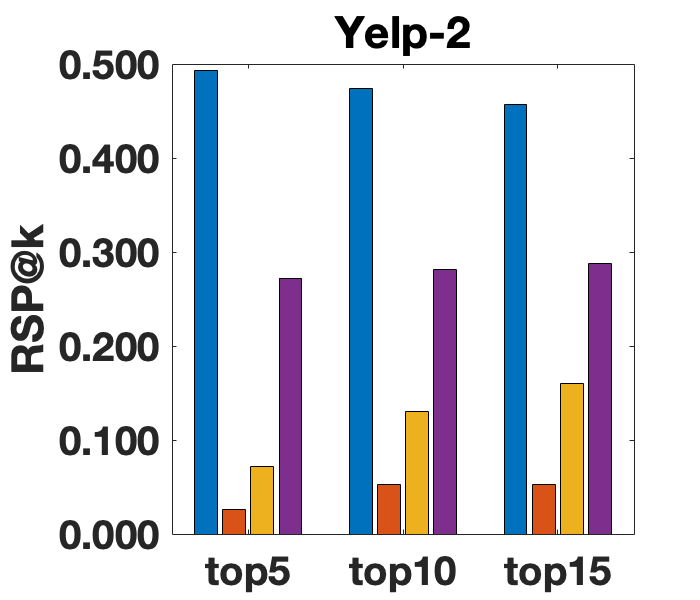}
        %\vspace{-7pt}
        \label{fig:RQ2_SP_SP_Yelp}
    \end{subfigure}
            ~ 
    \begin{subfigure}[t]{0.212\textwidth}
        \centering
        \includegraphics[width=1\textwidth]{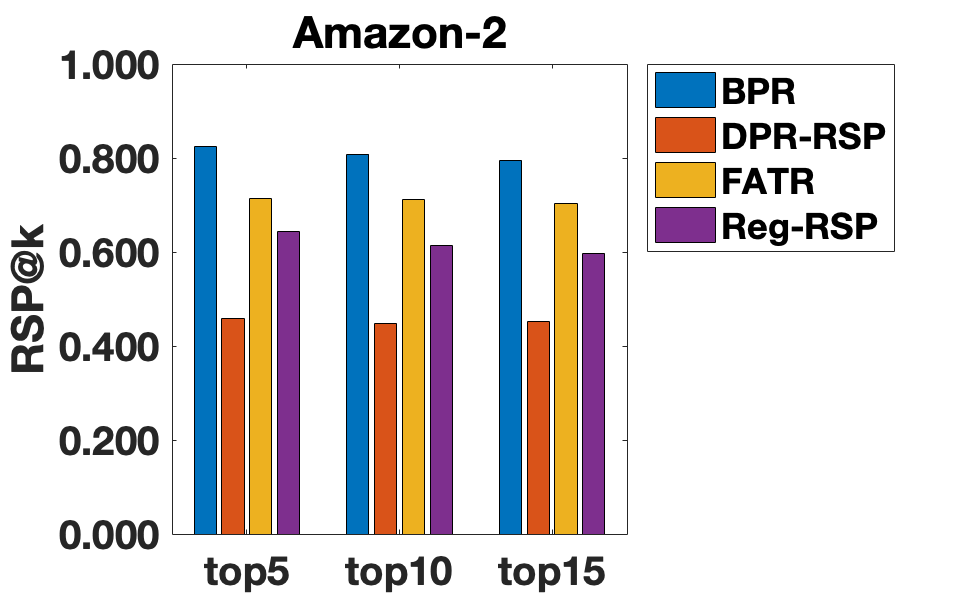}
        %\vspace{-7pt}
        \label{fig:RQ2_SP_SP_Amazon}
    \end{subfigure}
    \vspace{-20pt}
    \caption{F1@k and RSP@k of four different models over three datasets.}
    \label{fig:RQ2_SP} 
    % \vspace{-8pt}
\end{figure*}

\begin{figure*}[t!]
    \centering
    \begin{subfigure}[t]{0.155\textwidth}
        \centering
        \includegraphics[width=1\textwidth]{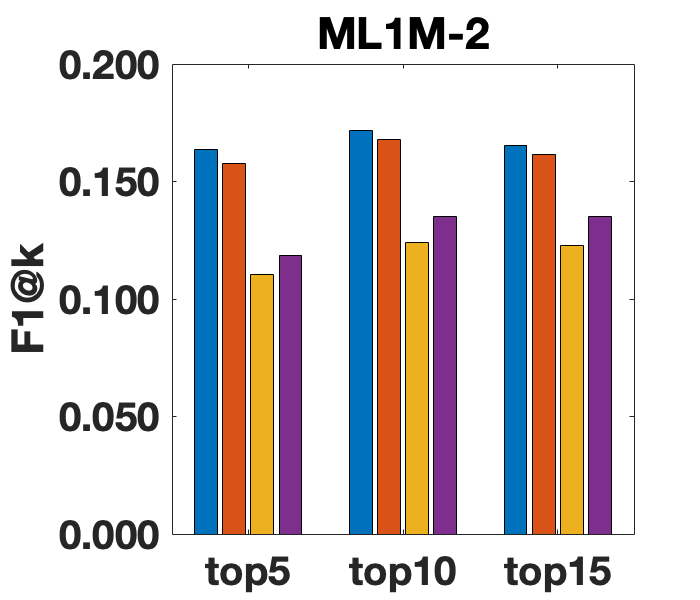}
        %\vspace{-7pt}
        \label{fig:RQ2_EO_F_ML1M}
    \end{subfigure}
        ~ 
    \begin{subfigure}[t]{0.155\textwidth}
        \centering
        \includegraphics[width=1\textwidth]{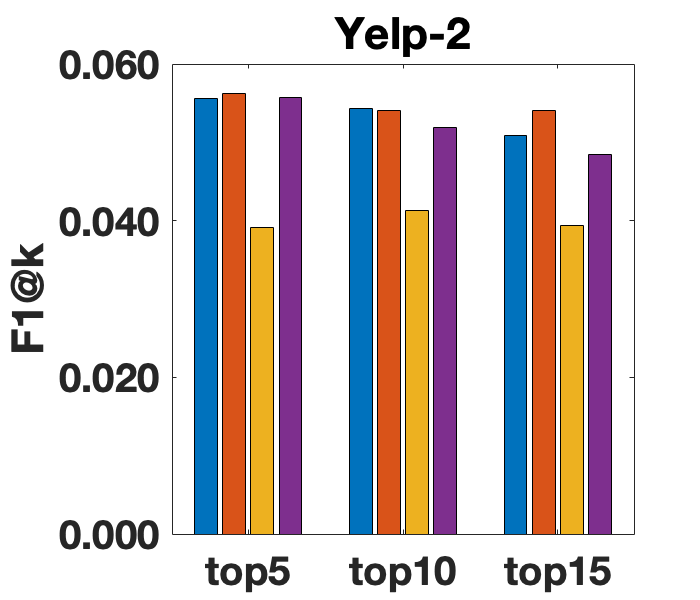}
        %\vspace{-7pt}
        \label{fig:RQ2_EO_F_Yelp}
    \end{subfigure}
            ~ 
    \begin{subfigure}[t]{0.155\textwidth}
        \centering
        \includegraphics[width=1\textwidth]{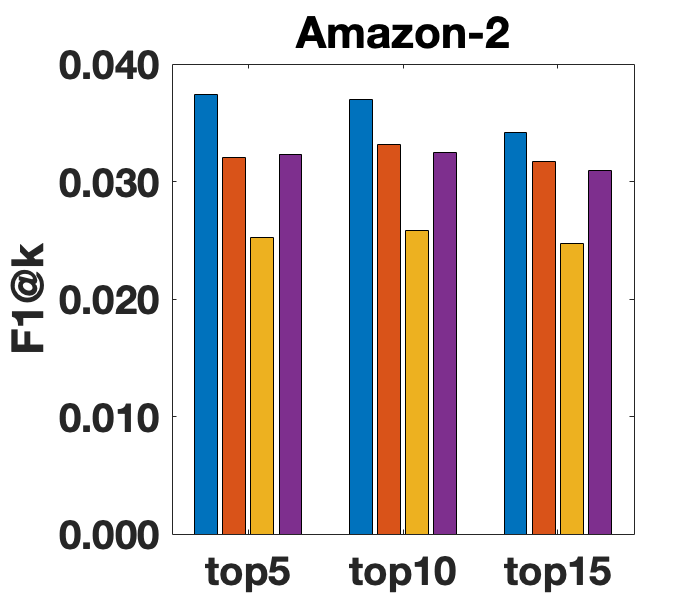}
        %\vspace{-7pt}
        \label{fig:RQ2_EO_F_Amazon}
    \end{subfigure}
            ~ 
    \begin{subfigure}[t]{0.155\textwidth}
        \centering
        \includegraphics[width=1\textwidth]{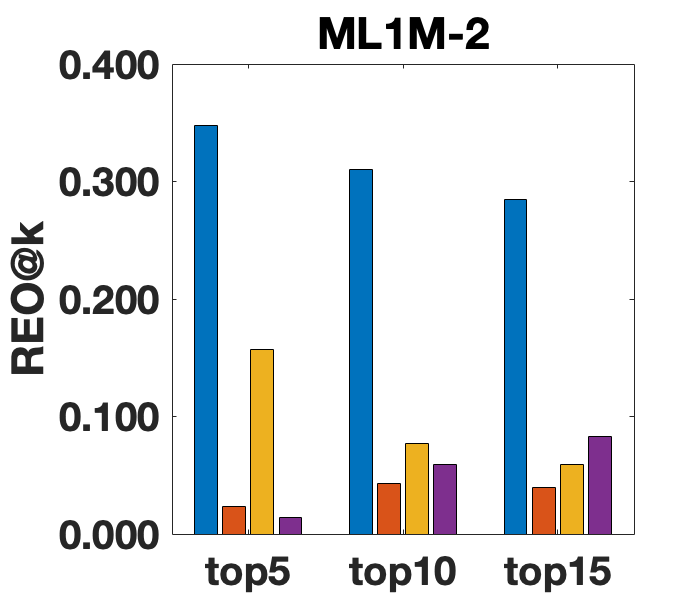}
        %\vspace{-7pt}
        \label{fig:RQ2_EO_EO_ML1M}
    \end{subfigure}
            ~ 
    \begin{subfigure}[t]{0.155\textwidth}
        \centering
        \includegraphics[width=1\textwidth]{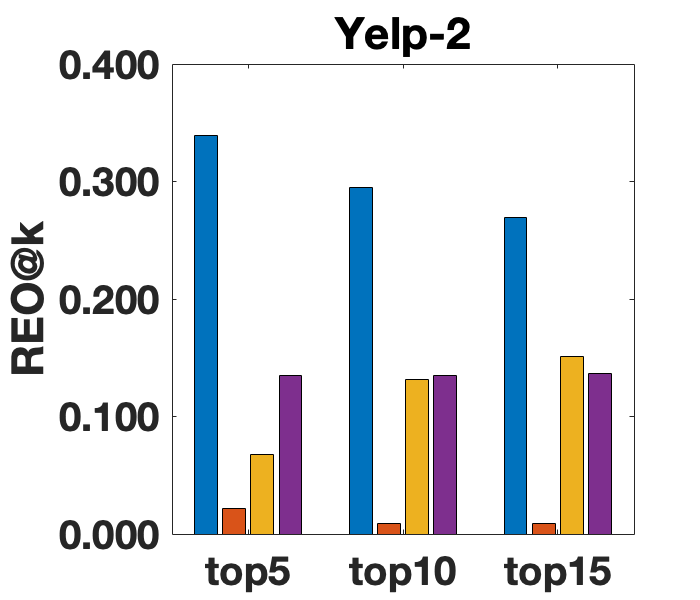}
        %\vspace{-7pt}
        \label{fig:RQ2_EO_EO_Yelp}
    \end{subfigure}
            ~ 
    \begin{subfigure}[t]{0.212\textwidth}
        \centering
        \includegraphics[width=1\textwidth]{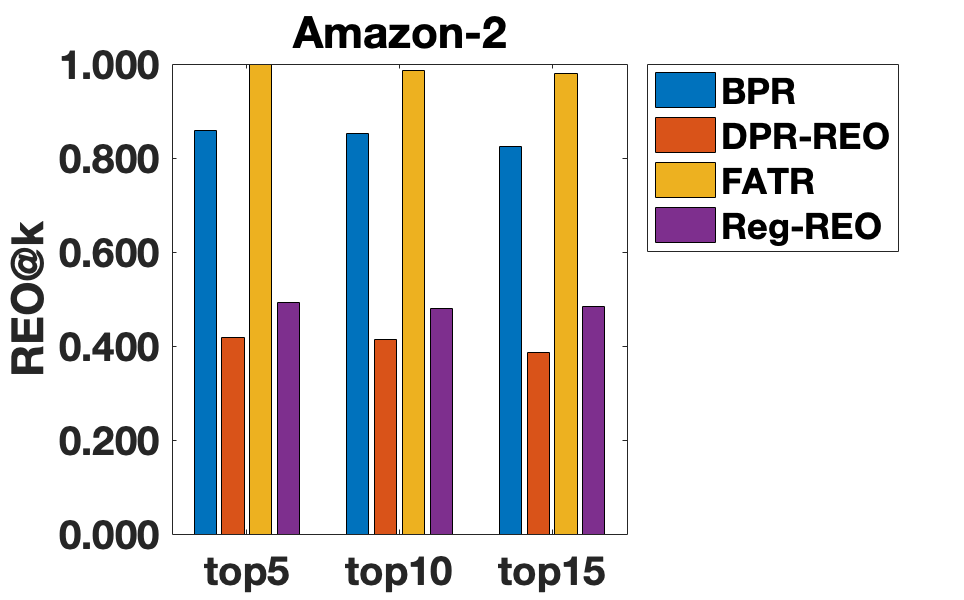}
        %\vspace{-7pt}
        \label{fig:RQ2_EO_EO_Amazon}
    \end{subfigure}
    \vspace{-20pt}
    \caption{F1@k and REO@k of four different models over three datasets.}
    \label{fig:RQ2_EO} 
    \vspace{-5pt}
\end{figure*}

We next compare the proposed DPR with state-of-the-art alternatives to answer two questions: (i) how does the proposed adversarial learning perform in comparison with baselines for predicted score distribution similarity enhancement? and (ii) how does the proposed DPR perform for both fairness metrics compared with baselines? Because baselines Reg-RSP and Reg-REO can only work for binary-group cases, we conduct the experiment over ML1M-2, Yelp-2, and Amazon-2 datasets in this subsection.

To answer the first question, we report JS Divergences of score distributions for different groups in Table~\ref{table:RQ2}, where the top five rows are calculated on all user-item pairs not in the training set (fitting the RSP setting), and the bottom five rows are calculated on user-item pairs only in the test set (fitting the REO setting). The improvement rates of DPR over the best baselines also are calculated. From the table we can conclude that the proposed adversarial learning can more effectively enhance score distribution similarity than baselines. Although less competitive, both FATR and Reg models can improve the distribution similarity to some degree compared with BPR.

As for the second question, we show $F1@k$, $RSP@k$, and $REO@k$ comparison between all methods over all datasets in Figure~\ref{fig:RQ2_SP} and Figure~\ref{fig:RQ2_EO}. On the one hand, from the leftmost three figures in both Figure~\ref{fig:RQ2_SP} and Figure~\ref{fig:RQ2_EO}, we can observe that DPR-RSP and DPR-REO preserve relatively high $F1@k$ from BPR and outperform other baselines significantly. On the other hand, from the rightmost three figures, we are able to see that DPR-RSP and DPR-REO enhance RSP and REO to a great extent respectively, which also outperform other fairness-aware methods considerably. Besides, one potential reason for better recommendation quality for DPR on Yelp is that the intrinsic bias in Yelp is small, thus DPR can promote unpopular groups and keep the original high rankings for popular groups simultaneously, leading to better recommendation performance.

\subsection{RQ3: Impact of Hyper-Parameters}

Finally, we investigate the impact of three hyper-parameters: (i) the number of layers in the MLP adversary; (ii) the adversary trade-off regularizer $\alpha$; and (iii) the KL-loss trade-off regularizer $\beta$. For conciseness, we only report experimental results on ML1M dataset, but note that the results on other datasets show similar patterns.

\begin{figure}[t!]
    \centering
    \begin{subfigure}[t]{0.22\textwidth}
        \centering
        \includegraphics[width=1\textwidth]{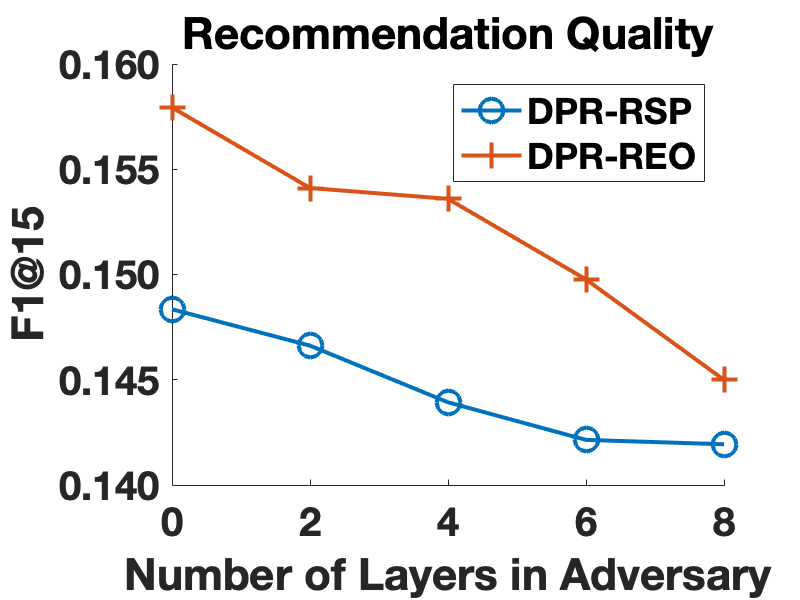}
        %\vspace{-7pt}
        \label{fig:RQ3_NN_Rec}
    \end{subfigure}%
        ~ %\hspace*{15pt}
    \begin{subfigure}[t]{0.22\textwidth}
        \centering
        \includegraphics[width=1\textwidth]{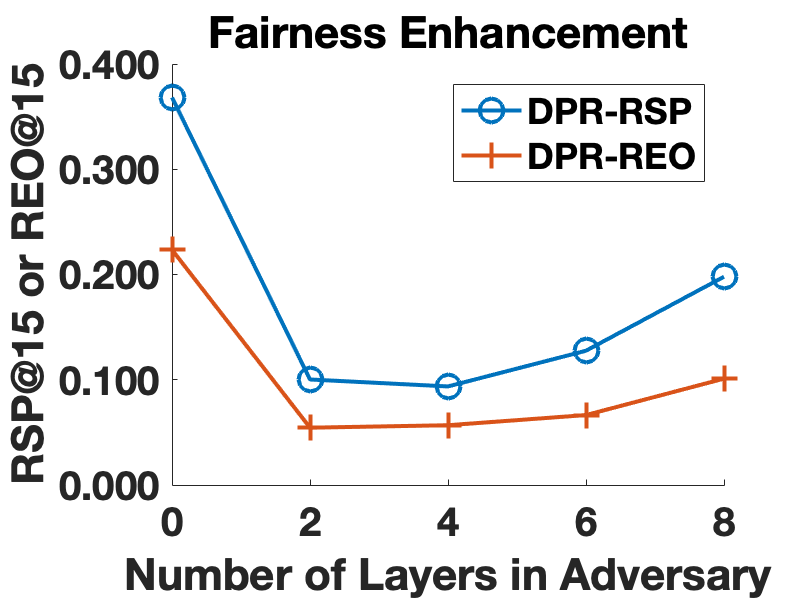}
        %\vspace{-7pt}
        \label{fig:RQ3_NN_Fair}
    \end{subfigure}
    \vspace{-15pt}
    \caption{$F1@15$, $RSP@15$, and $REO@15$ of DPR-RSP and DPR-REO w.r.t. different numbers of layers over ML1M.}
    \label{fig:RQ3_NN} 
    \vspace{-15pt}
\end{figure}

\smallskip
\noindent\textbf{Impact of Layers in Adversary.} First, we experiment with the number of layers in MLP adversary varying in $\{0, 2, 4, 6, 8\}$, and the other parameters are the same as introduced in Section~\ref{sec:setup} including that the number of neurons in each MLP layer is still 50. Generally speaking, with more layers, the adversary is more complex and expressive, which intuitively results in better fairness performance. The $F1@15$ results of DPR-RSP and DPR-REO w.r.t. different numbers of layers are shown at the left in Figure~\ref{fig:RQ3_NN}, and $RSP@15$ and $REO@15$ results are presented at the right in Figure~\ref{fig:RQ3_NN}. From these figures, we can infer that with a more powerful adversary, the recommendation quality drops more; however, the fairness improvement effect first gets promoted but then weakened due to difficulty of model training. The best value is around 2 to 4. Besides, we can also find that it is easier to augment the metric REO than RSP with less recommendation quality sacrificed, which is consistent with the observation in Section~\ref{sec:RQ1}.

\begin{figure}[t!]
    \centering
    \begin{subfigure}[t]{0.22\textwidth}
        \centering
        \includegraphics[width=1\textwidth]{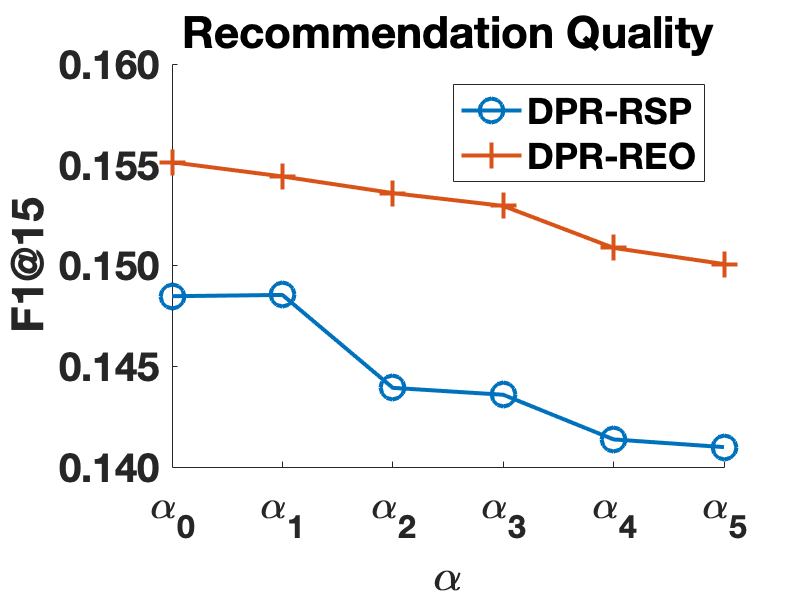}
        %\vspace{-7pt}
        \label{fig:RQ3_alpha_Rec}
    \end{subfigure}%
        ~ %\hspace*{15pt}
    \begin{subfigure}[t]{0.22\textwidth}
        \centering
        \includegraphics[width=1\textwidth]{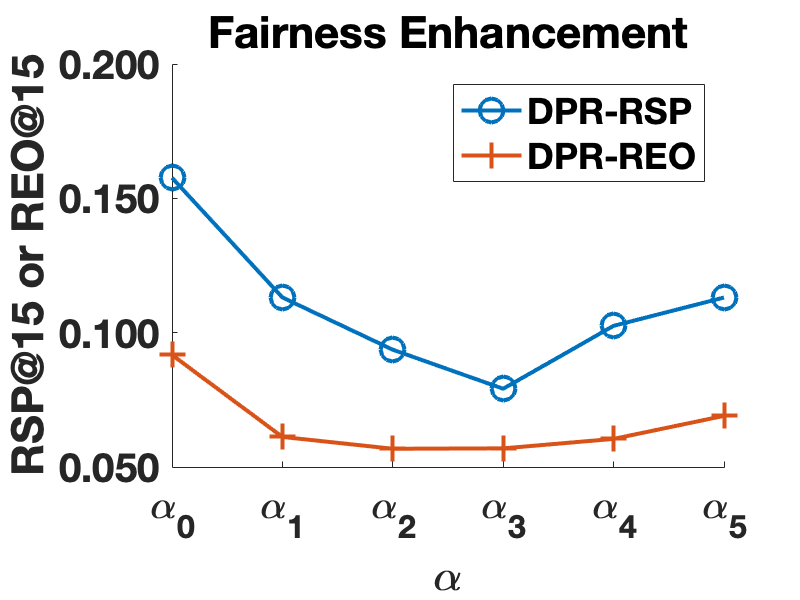}
        %\vspace{-7pt}
        \label{fig:RQ3_alpha_Fair}
    \end{subfigure}
    \vspace{-20pt}
    \caption{$F1@15$, $RSP@15$ and $REO@15$ of DPR-RSP and DPR-REO w.r.t. different $\alpha$ over ML1M.}
    \label{fig:RQ3_alpha} 
    \vspace{-10pt}
\end{figure}

\smallskip
\noindent\textbf{Impact of $\alpha$.} Then, we vary the adversary trade-off regularizer $\alpha$ and plot the results in Figure~\ref{fig:RQ3_alpha}, where the x-axis coordinates $\{\alpha_0,\alpha_1,\ldots,\alpha_5\}$ are $\{1000, 3000, 5000, 7000, 9000, 11000\}$ for DPR-RSP and $\{200, 600,$ $ 1000, 1400, 1800, 2200\}$ for DPR-REO. The left figure demonstrates the $F1@15$ results with different $\alpha$, which shows that with larger weight for the adversary, the recommendation quality decreases more. For the fairness improving performance, as presented at the right in Figure~\ref{fig:RQ3_alpha}, with larger $\alpha$, both DPR-RSP and DPR-REO first improve the fairness, but then increase it again, which is most likely due to the dominating of adversary over KL-loss in the objective function. To balance the recommendation quality and recommendation fairness, setting $\alpha=5000$ for DPR-RSP and $\alpha=1000$ for DPR-REO are reasonable choices.

\begin{figure}[t!]
    \centering
    \begin{subfigure}[t]{0.22\textwidth}
        \centering
        \includegraphics[width=1\textwidth]{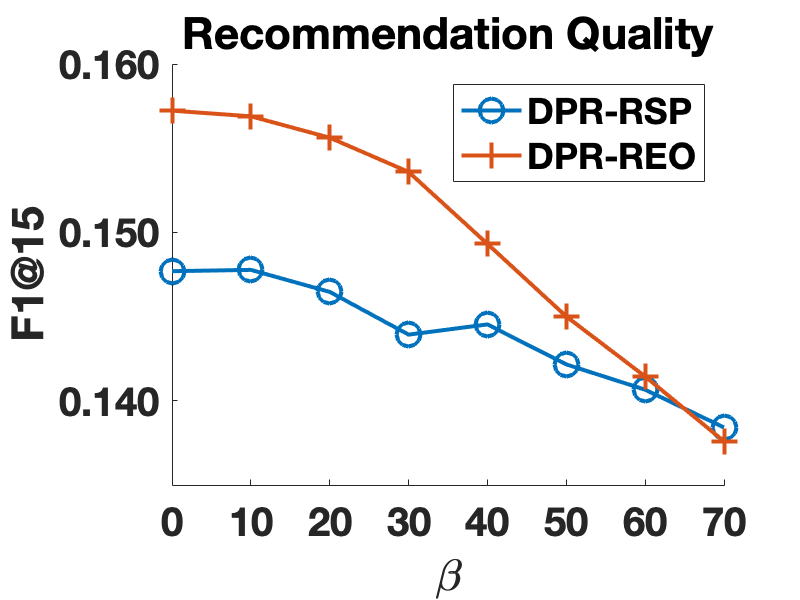}
        %\vspace{-7pt}
        \label{fig:RQ3_beta_Rec}
    \end{subfigure}%
        ~ %\hspace*{15pt}
    \begin{subfigure}[t]{0.22\textwidth}
        \centering
        \includegraphics[width=1\textwidth]{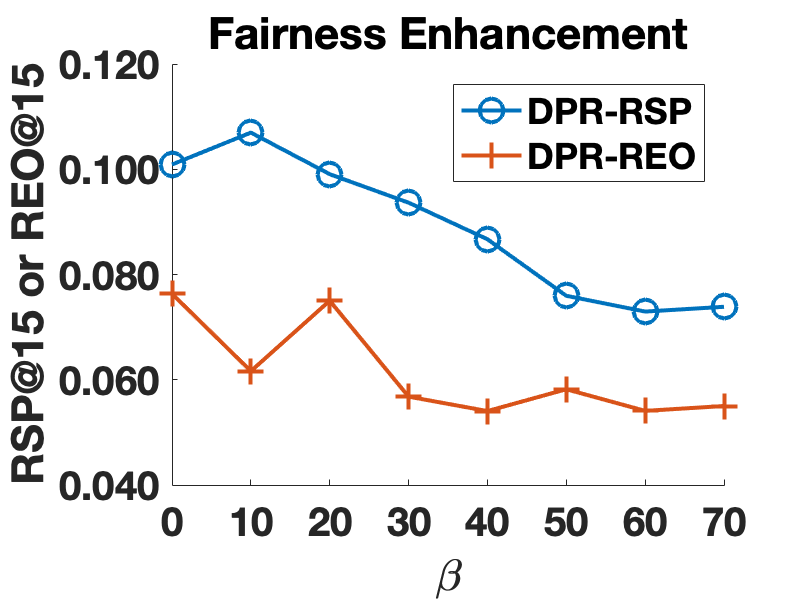}
        %\vspace{-7pt}
        \label{fig:RQ3_beta_Fair}
    \end{subfigure}
    \vspace{-20pt}
    \caption{$F1@15$, $RSP@15$ and $REO@15$ of DPR-RSP and DPR-REO w.r.t. different $\beta$ over ML1M.}
    \label{fig:RQ3_beta} 
    \vspace{-10pt}
\end{figure}

\smallskip
\noindent\textbf{Impact of $\beta$.} Last, we study the impact of the KL-loss trade-off regularizer $\beta$ and vary the value in the set $\{0, 10, 20, 30 , 40, 50, 60, 70\}$. The left figure in Figure~\ref{fig:RQ3_beta} shows the change tendency of $F1@15$, which implies that larger $\beta$ leads to lower recommendation quality. The fairness improving performance of DPR-RSP and DPR-REO with different $\beta$ are shown at the right in Figure~\ref{fig:RQ3_beta}, from which we can observe that with higher $\beta$, the fairness is enhanced better, and converges to a certain degree. However, the impact of $\beta$ is not as strong as that of $\alpha$ (the value changes of $RSP@15$, and $REO@15$ in Figure~\ref{fig:RQ3_beta} are smaller than those in Figure~\ref{fig:RQ3_alpha}).

\section{Conclusion}
In this paper, based on well known concepts of statistical parity and equal opportunity, we first propose two fairness metrics designed specifically for personalized ranking recommendation tasks. Then we empirically show that the influential Bayesian Personalized Ranking model is vulnerable to the inherent data imbalance and tends to generate unfair recommendations w.r.t. the proposed fairness metrics. Next we propose a novel fairness-aware personalized ranking model incorporating adversarial learning to augment the proposed fairness metrics. At last, extensive experiments show the effectiveness of the proposed model over other state-of-the-art alternatives. 